\documentclass[letterpaper,11pt]{article}
\usepackage[letterpaper,hmargin=1in,vmargin=1.25in]{geometry}
\usepackage{amsmath}                    % AMS math
\usepackage{amssymb}                    % special AMS math symbols
\usepackage{amsthm}                     % AMS theorem/proof
\usepackage{amsfonts}                   % more special AMS math symbols
\usepackage{url}                        % \url{...} formatting URLs
\usepackage{cite}                       % sort bibliographical entries
\usepackage{hyperref}                   % add hyper-references
\hypersetup{colorlinks=true,linkcolor=[rgb]{0,0,0},citecolor=[rgb]{0,0,0},urlcolor=[rgb]{0,0,0}}
\usepackage{enumitem}                   % added controls for enumerations
\usepackage{graphicx}                   % graphics (includegraphics)
\usepackage{xcolor}
\definecolor{myfuchsia}{HTML}{FF00FF}
\usepackage{accents}                    % for \dbtilde

\newtheorem{thm}{Theorem}

\newtheorem{lem}{Lemma}[section]

\newtheorem{definition}{Definition}[section]
\newcommand{\ang}[1]{\langle #1\rangle}
\newcommand{\RE}{\mathbb{R}}  % real space
\newcommand{\eps}{\varepsilon}  % my preferred epsilon
\newcommand{\etal}{\textit{et~al.}} % ``et al.''
\newcommand{\ST}{\,:\,}  % { x \ST y }

\newcommand{\inv}[1]{\frac{1}{#1}}
\newcommand{\Gradient}{\nabla}

\newcommand{\bd}{\partial}
\newcommand{\subgrad}{\partial} % Different term, same symbol
\newcommand{\inner}[2]{\ang{#1,#2}} % inner product
\newcommand{\stdpolar}[1]{{#1}^{\circ}} % standard polar
\newcommand{\SP}{\kern+1pt}
\DeclareMathOperator{\diam}{diam}
\DeclareMathOperator{\vol}{vol}
\DeclareMathOperator{\area}{area}
\DeclareMathOperator{\vrad}{vrad}
\DeclareMathOperator{\arad}{arad}

\DeclareMathOperator{\interior}{int}

\DeclareMathOperator{\dom}{dom}
\DeclareMathOperator{\graph}{graph}
\DeclareMathOperator{\epi}{epi}
\DeclareMathOperator{\VCap}{VCap} % volume cap
\DeclareMathOperator{\SCap}{SCap} % surface cap
\DeclareMathOperator{\DCap}{DCap} % dual cap
\DeclareMathOperator{\SCapD}{SCap^{\kern-1pt *}} % surface cap in the dual
\DeclareMathOperator{\VCapD}{VCap^{\kern-1pt *}} % volume cap in the dual
\DeclareMathOperator{\DBase}{DBase} % base of a dual cap
\DeclareMathOperator{\VBase}{VBase} % base of a volume cap
\DeclareMathOperator{\Cone}{Cone} % cone
\newcommand{\II}{I^{\pm}} % signed index set
\newcommand{\dbtilde}[1]{\accentset{\approx}{#1}}
\newcommand{\simtilde}[1]{\accentset{\sim}{#1}}

\begin{document}
%-----------------------------------------------------------------------

\title{Optimal Area-Sensitive Bounds for Polytope Approximation\thanks{A preliminary version of this paper appeared in the 28th Symp.\ on Computational Geometry, 2012.}}

\author{%
	Sunil Arya\thanks{Research supported by the Research Grants Council of Hong Kong, China under project numbers 610012 and 16213219.}\\
		Department of Computer Science and Engineering \\
		Hong Kong University of Science and Technology \\
		Clear Water Bay, Kowloon, Hong Kong\\
		arya@cse.ust.hk \\
		\and
	Guilherme D. da Fonseca\thanks{Supported by the French ANR PRC grant ADDS (ANR-19-CE48-0005).}\\
            Aix-Marseille Universit\'{e} and LIS \\
            Marseille, France \\
            guilherme.fonseca@lis-lab.fr \\
            \and
	David M. Mount\thanks{Research supported by NSF grant CCF-1618866.}\\
		Department of Computer Science and \\
		Institute for Advanced Computer Studies \\
		University of Maryland \\
		College Park, Maryland 20742 \\
		mount@cs.umd.edu \\
}

\date{}

\maketitle

%=======================================================================
% Abstract
%=======================================================================
\begin{abstract}
Approximating convex bodies is a fundamental problem in geometry. Given a convex body $K$ in $\RE^d$ for a fixed dimension $d$, the objective is to minimize the number of facets of an approximating polytope for a given Hausdorff error $\eps$. The best known uniform bound, due to Dudley (1974), shows that $O((\diam(K)/\eps)^{(d-1)/2})$ facets suffice. Although this bound is optimal for fat objects, such as Euclidean balls, it is far from optimal for ``skinny'' convex bodies.

Skinniness can be characterized relative to the Euclidean ball. Given a convex body $K$, define its \emph{area radius}, $\arad(K)$, to be the radius of the Euclidean ball having the same surface area as $K$. It follows from generalizations of the isoperimetric inequality that $\diam(K) \geq 2 \cdot \arad(K)$.

We show that, given a convex body whose minimum width is at least $\eps$, it is possible to approximate the body by a polytope having $O((\arad(K)/\eps)^{(d-1)/2})$ facets. Our approach works by first reducing the problem of approximating convex bodies to that of approximating convex functions. We employ a classical concept from convexity, called Macbeath regions. We demonstrate that there is a polar relationship between the Macbeath regions of a function and the Macbeath regions of its Legendre dual. This is combined with known bounds on the Mahler volume to bound the total size of the approximation.
\end{abstract}

\newpage
%=======================================================================
\section{Introduction} \label{sec:intro}
%=======================================================================

Approximating convex bodies by polytopes is a well-studied problem in computational and combinatorial geometry. (See Bronstein~\cite{Bro08} for a survey.) Given a convex body $K$ in the Euclidean $d$-dimensional space and a scalar $\eps > 0$, we say that a polytope $P$ is an \emph{$\eps$-approximation} to $K$ if the Hausdorff distance between $K$ and $P$ is at most $\eps$. (See Section~\ref{sec:notation} for definitions and notation.) The question we consider is how many facets are needed in such a polytope. Throughout, we assume that the dimension $d$ is a constant, and our asymptotic forms conceal constant factors that depend on $d$.

Approximation bounds from the literature come in two common forms. In both cases, the bounds hold for all $\eps \leq \eps_0$, for some given $\eps_0$. Bounds are said to be \emph{nonuniform} if the value of $\eps_0$ depends on properties of $K$. Nonuniform bounds often hold subject to smoothness conditions on $K$'s boundary (e.g., $K$'s boundary is $C^2$ continuous). Examples include the works of Gruber~\cite{Gru93a}, Clarkson~\cite{Cla06}, and others~\cite{Bor00,Sch87,McV75,Tot48}. In contrast, in uniform bounds, the value of $\eps_0$ is independent of $K$, but can depend on $d$. Such bounds hold without any additional smoothness assumptions. Examples include the results of Dudley~\cite{Dud74} and Bronshteyn and Ivanov~\cite{BrI76}. Our results are of this latter type.

Dudley~\cite{Dud74} showed that any convex body $K$ can be $\eps$-approximated by a polytope $P$ with at most $c_d \cdot (\diam(K)/\eps)^{(d-1)/2}$ facets, where $c_d$ is a constant depending on the dimension, $\diam(K)$ denotes $K$'s diameter, and $0 < \eps \leq \diam(K)$. Bronshteyn and Ivanov showed that the same asymptotic bound holds for the number of vertices. Up to constant factors depending on the dimension, both results are known to be tight in the worst case.

The bounds given by both Dudley and Bronshteyn--Ivanov are tight in the worst case up to constant factors~\cite{Bro08}, with the worst case arising when $K$ is a Euclidean ball. These bounds may be significantly suboptimal if $K$ is ``skinny''. The skinniness of a convex body can be measured relative to a Euclidean ball. Let $B^d_2$ denote the Euclidean unit ball in $\RE^d$. Define the \emph{volume radius} of a convex body $K$ in $\RE^d$, denoted $\vrad(K)$, to be the radius of the Euclidean ball of the same volume as $K$, that is,
\[
    \vrad(K)
        ~ = ~ \left( \frac{\vol_d(K)}{\vol_d(B^d_2)} \right)^{\kern-2pt 1/d},
\]
where $\vol_k(\cdot)$ denotes the $k$-dimensional Lebesgue measure. We can similarly define the \emph{area radius}, denoted $\arad(K)$, in terms of $K$'s surface area as
\[
    \arad(K)
        ~ = ~ \left( \frac{\area(K)}{\area(B^d_2)} \right)^{\kern-2pt \frac{1}{d-1}},
\]
where $\area(K) = \vol_{d-1}(\bd K)$. These quantities are closely related to the classical concepts of \emph{quermassintegrals} and of \emph{intrinsic volumes}~\cite{McM75,McM91}. From generalizations of the isoperimetric inequality it follows that $\vrad(K) \leq \arad(K) \leq \diam(K)/2$ (see, e.g., \cite{McM91}).

In this paper, we strengthen Dudley's bound by showing that the complexity of approximation can be made sensitive to $K$'s skinniness, as expressed in terms of its area radius. Here is our main result.

%-----------------------------------------------------------------------
\begin{thm} \label{thm:main}
Consider any convex body $K$ in $\RE^d$ and any $\eps > 0$ such that the width of $K$ in any direction is at least $\eps$. There exists an outer $\eps$-approximating polytope $P$ for $K$ whose number of facets is at most
\[
    c_d \left(\frac{\arad(K)}{\eps}\right)^{\kern-2pt \frac{d-1}{2}},
\]
where $c_d$ is a constant (depending on $d$).
\end{thm}
%-----------------------------------------------------------------------

By \emph{outer approximation}, we mean that $P \supseteq K$. As a function of surface area or area radius alone, the bound of Theorem~\ref{thm:main} is tight up to constant factors. To see why, observe that the bound can be stated in terms of $K$'s surface area as $c_d \sqrt{\area(K)} / \eps^{(d-1)/2}$. By the isoperimetric inequality, up to constant factors, $\area(K) \leq \diam(K)^{d-1}$, and the tightness of Dudley's bound implies that the number of facets needed is at least $(\diam(K)/\eps)^{(d-1)/2} \geq \sqrt{\area(K)} / \eps^{(d-1)/2}$.

Any bound that holds in the uniform setting also holds in the nonuniform setting. Of course, in the absence of the uniformity requirement, a simpler analysis may be possible. In Section~\ref{sec:nonuniform} we present a simple derivation of a reformulation of Theorem~\ref{thm:main} in the nonuniform setting.

As an additional contribution of this paper, we show that convex-body approximation in $\RE^d$ can be reduced to convex-function approximation in $\RE^{d-1}$ (see Lemma~\ref{lem:outer} in Section~\ref{sec:support-sets}). Given a lower semicontinuous convex function $f: \RE^{d-1} \to \RE \cup \{+\infty\}$ and $D \subseteq \RE^{d-1}$, we say that a lower semicontinuous convex function $\widehat{f}$ is a \emph{convex lower $\eps$-approximation to $f$ on $D$} if
\begin{itemize}
\item $\widehat{f}(x) \leq f(x)$, for all $x \in \RE^{d-1}$ and 
\item $\widehat{f}(x) \geq f(x) - \eps$, for all $x \in D$. 
\end{itemize}
(Observe that $\widehat{f}$ is pointwise on or below $f$ throughout $\RE^{d-1}$, and it is $\eps$-close throughout $D$.)
Our main result on functional approximation is presented below. (See Section~\ref{sec:notation} for definitions.)

%-----------------------------------------------------------------------
\begin{thm}[Area-Sensitive Functional Approximation] \label{thm:func-approx}
Let $D$ be a compact convex domain in $\RE^{d-1}$ of minimal width at least $\eps$, and let $f: \RE^{d-1} \to \RE \cup \{+\infty\}$ be a lower semicontinuous convex function that is Lipschitz continuous over $D$ with Lipschitz constant $\lambda$. Then, for some constant $c_d$ (depending on $d$) there exists a set of at most
\[
    c_d \left( \max(1, \lambda) \cdot \frac{\vrad(D)}{\eps} \right)^{\kern-2pt \frac{d-1}{2}}
\]
affine functions on $\RE^{d-1}$ such that the pointwise maximum of these functions is a convex lower $\eps$-approximation to $f$ on $D$.
\end{thm}
%-----------------------------------------------------------------------

Note that since $D$ resides in $\RE^{d-1}$, its volume radius in the above theorem is defined with respect to the $(d-1)$-Lebesgue measure.

%=======================================================================
\subsection{Related Work} \label{sec:related}
%=======================================================================

The problem of shape-sensitive approximations was considered by Bonnet~\cite{Bon18}, who studied the problem in the uniform setting (which he calls the non-asymptotic, non-smooth case). His results are presented in terms of the intrinsic volume $V_i(K)$, for $1 \leq i \leq d$. If we define $\text{rad}_i(K)$ to be $V_i(K)^{1/i}$, his results imply that there exists an $\eps$-approximation to $K$ with $O((\diam(K) \delta^{\beta}/\eps)^{(d-1)/2})$ facets, where $\delta = \text{rad}_{(d-1)/2}(K) / \diam(K)$ and $\beta$ is roughly $1/(2d)$. He conjectures that the results hold for $\beta = 1$ and with $\text{rad}_{d-1}(K) = \arad(K)$ in place of $\text{rad}_{(d-1)/2}(K)$. Since, up to constant factors, $\text{rad}_{d-1}(K) = \arad(K)$, this essentially matches the bound of Theorem~\ref{thm:main}. We also considered the problem of an area-sensitive approximation in an earlier work~\cite{AFM12a}. The bound presented there was worse by a factor of $\log (1/\eps)$. 

The univariate case of convex function approximation was studied by Rote~\cite{Rot92}. His bounds match ours for the $d=1$ case. In a recent paper, we presented a volume-sensitive bound by proving the existence of an approximation with $O((\vrad(K)/\eps)^{(d-1)/2})$ facets~\cite{ArM25}. As observed above, $\arad(K) \geq \vrad(K)$, and hence that bound subsumes the area-sensitive bound for the case of convex bodies. The results presented here are still of interest for a couple of reasons. Our results on convex function approximation are novel, and the functional perspective on approximation yields new insights. For example, in Section~\ref{sec:cap-dual-cap} (Lemma~\ref{lem:dual-bases}) we show how natural cap-like structures of a convex function and its dual conjugate are related through polarity in the domain space. Another feature of our area-based bounds is that they can be applied to surface patches of convex bodies. There are applications involving unbounded objects, such as in the generation of space-efficient minimization diagrams for approximate nearest-neighbor searching~\cite{HaK15,AAFM19}, where area sensitivity is meaningful, but volume sensitivity is not. 

%=======================================================================
\subsection{Notation and Background} \label{sec:notation}
%=======================================================================

Throughout, $K$ denotes a convex body in $\RE^d$, that is, a compact convex subset with a nonempty interior, and $\eps$ denotes a fixed approximation parameter. Let $\bd K$ denote the boundary of $K$. Let $\vol(K) = \vol_d(K)$ denote its $d$-dimensional Lebesgue measure, and let $\area(K) = \vol_{d-1}(\bd K)$ denote its surface area. For $\alpha \geq 0$, $\alpha K$ denotes a uniform scaling of $K$ about the origin, and for $x \in \RE^{d}$, $K + x$ denotes the translation of $K$ by $x$. Given a convex body $L$, let $K \oplus L$ denote the Minkowski sum of $K$ and $L$, that is, $\{x + y \ST x \in K, y \in L\}$. Let $B^d_2$ denote the Euclidean ball of unit radius centered at the origin.

Throughout, we use $\inner{\cdot}{\cdot}$ to denote the standard inner (dot) product and use $\|\cdot\| = \sqrt{\inner{\cdot}{\cdot}}$ to denote the Euclidean norm. Given two convex bodies $K$ and $L$ in $\RE^d$, their \emph{Hausdorff distance} is defined to be
\[
    \min \left\{ r \geq 0 \ST K \subseteq L \oplus r B^d_2 ~\text{and}~ L \subseteq K \oplus r B^d_2 \right\}.
\]
Given a unit vector $u$, the \emph{width of $K$ in direction $u$} is the smallest distance between two hyperplanes, both orthogonal to $u$, that enclose $K$. The \emph{minimum width} of $K$ is the minimum over all directional widths.

Next, we review some standard concepts from convex analysis (see, e.g., Rockafellar~\cite{Roc97}). Consider a lower semicontinuous convex function $f$, where $f: \RE^{d-1} \to \RE \cup \{+\infty\}$. Let $\dom f$ denote its \emph{effective domain}, that is, the set of $x \in \RE^{d-1}$ such that $f(x)$ is finite. To relate sets in $\RE^d$ and functions on $\RE^{d-1}$, we will often express points in $\RE^d$ as a coordinate pair $(x;t) \in \RE^{d-1} \times \RE$. Each such point naturally defines an associated hyperplane
\[
    h_{(x; t)}
        ~ = ~ \left\{ (p; \tau) \in \RE^d \ST \tau = \inner{p}{x} - t \right\}.
\]
Clearly, $(p; \tau) \in h_{(x;t)}$ if and only if $(x;t) \in h_{(p; \tau)}$. Such a hyperplane defines an \emph{upper halfspace} ($\tau \geq \inner{p}{x} - t$) and \emph{lower halfspace} ($\tau \leq \inner{p}{x} - t$). The \emph{graph} of $f$, denoted $\graph f$, is $\{(x;t) \in \RE^{d-1} \times \RE \ST t = f(x)\}$. Its \emph{epigraph}, denoted $\epi f$, is defined similarly, but where $t \geq f(x)$. The epigraph of a convex function is a convex set.

Given a lower semicontinuous convex function $f$ and $x_0 \in \interior(\dom f)$, a vector $p \in \RE^{d-1}$ is a  \emph{subgradient} of $f$ at $x_0$ if
\[
    f(x)
        ~ \geq ~ f(x_0) + \inner{p}{x - x_0}, ~~\text{for all $x \in \RE^{d-1}$}.
\]
This generalizes the notion of gradient to non-smooth functions. The following is well known (see, e.g.,~\cite{Roc67}). 

%-----------------------------------------------------------------------
\begin{lem} \label{lem:subgradient}
Given a lower semicontinuous convex function $f$ and $x \in \interior(\dom f)$, a functional $p$ is a subgradient of $f$ at $x$ if and only if $(p; -1)$ is the outer normal vector of a hyperplane supporting the epigraph of $f$ at $(x; f(x))$.
\end{lem}
%-----------------------------------------------------------------------

The set of all subgradients at $x_0$ is called the \emph{subdifferential} of $f$ at $x_0$, denoted $\subgrad f(x_0)$.  A function $f$ is \emph{smooth} if it has a well-defined \emph{gradient} at each point $x_0 \in \interior(\dom f)$,
\[
    \Gradient f(x_0)
        ~ = ~ \left( \frac{\partial f(x_0)}{\partial x_1}, \ldots, \frac{\partial f(x_0)}{\partial x_{d-1}} \right).
\]
For such functions, $\subgrad f(x_0)$ consists of the single vector $\Gradient f(x_0)$. A convex function $f$ on a convex domain is \emph{strictly convex} if for all distinct $x, x' \in \dom f$
\[
    f((1-\gamma) x + \gamma x')
        ~ < ~ (1 - \gamma) f(x) + \gamma f(x'), ~~0 < \gamma < 1.
\]

A function $f$ is \emph{Lipschitz continuous} on a subset $D$ of its domain with constant $\lambda$, if $|f(x)-f(x')| \leq \lambda \|x-x'|$, for all $x, x' \in D$. The following is a straightforward consequence of this definition (see, e.g.,~\cite{Roc67}).

%-----------------------------------------------------------------------
\begin{lem} \label{lem:lipschitz}
Given a compact convex domain $D \subseteq \interior(\dom f)$, a convex function $f$ is Lipschitz continuous on $D$ with constant $\lambda$ if and only if $\|p\| \leq \lambda$, for every subgradient $p$ of $f$ at every point $x$ of $D$.
\end{lem}
%-----------------------------------------------------------------------

%=======================================================================
\subsection{Overview of Methods}
%=======================================================================

\begin{figure}[htbp]
    \centerline{\includegraphics[scale=0.4]{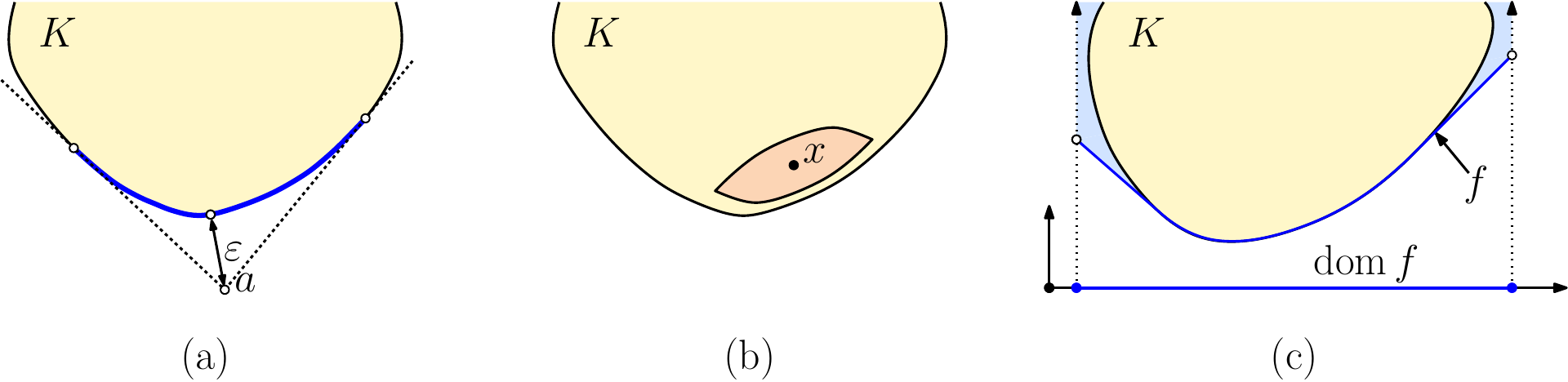}}
    \caption{Overview of methods: (a) dual cap, (b) Macbeath region, and (c) convex function induced by $K$.}
    \label{fig:methods}
\end{figure}

It is well known that computing a Hausdorff approximation to a convex body $K$ by a polytope can be reduced to sampling an appropriate set of points on the boundary of $K$ (see, e.g., \cite{BrG95,BrI76,Cla93}). For example, an outer approximation can be obtained by first sampling a sufficiently dense set of points on $K$'s boundary and then intersecting the halfspaces defined by the supporting hyperplanes at each of these points. The characterization of ``sufficiently dense'' can be based on the concept of hitting sets. Given a fixed approximation bound $\eps$, consider a point $a$ that is at distance $\eps$ from $K$ (see Figure~\ref{fig:methods}(a)). The set of points on the boundary of $K$ that are ``visible'' to $a$ is called a \emph{dual cap}. (Dual caps will be defined formally in Section~\ref{sec:cap-dual-cap}.) We say that a set $\mathcal{H}$ of points on $K$'s boundary is a \emph{hitting set} for dual caps if every dual cap contains at least one point of $\mathcal{H}$. It is straightforward to show that the polytope formed by the intersection of supporting halfspaces of $K$ at these points yields an $\eps$-approximation to $K$ in the Hausdorff sense.

%-----------------------------------------------------------------------
%-----------------------------------------------------------------------

A standard approach to constructing hitting sets is based on the concept of \emph{Macbeath regions}. This is a classical concept from the theory of convexity~\cite{Mac52}. Given a convex body $K$ and a point $x \in \interior(K)$, a Macbeath region is a centrally symmetric body centered at $x$, which adheres locally to $K$'s boundary (see Figure~\ref{fig:methods}(b)). A hitting set can be constructed by first generating a maximal set of disjoint Macbeath regions that lie within distance roughly $\eps$ of $K$'s boundary, then sampling points from the neighborhoods around each of these regions, and finally projecting these points onto $K$'s boundary. Unfortunately, a direct application of this process does not yield a good bound on the complexity of the approximation. The number of halfspaces in the approximation grows linearly with the number of Macbeath regions. Although a packing argument can be applied to bound the number of Macbeath regions of large volume, it is not easy to bound the number of Macbeath regions of small volume. 

In this paper, we introduce a novel approach to dealing with this issue. Our approach reduces the problem of approximating convex bodies to that of approximating convex functions. First, we show that it is possible to define $2 d$ lower semicontinuous convex functions, each over $\RE^{d-1}$, such that it is possible to combine individual (local) approximations to each of these functions to produce a single (global) approximation to $K$ (see Lemma~\ref{lem:outer}). To achieve area sensitivity, each function is Lipschitz continuous, and the area of its effective domain is not significantly larger than the surface area of $K$ (see Figure~\ref{fig:methods}(c)). 

Each function $f$ has the additional feature of being of Legendre type. This means that it has a well-defined Legendre transform $f^*$, and therefore there is a 1--1 correspondence between the effective domains of $f$ and $f^*$. This makes it possible to exploit the duality properties between these two functions. In particular, we show that there is a polar relationship between certain cap-like structures in the domains of $f$ and $f^*$ (see Lemma~\ref{lem:dual-bases}). We use known results on the Mahler volume to show that if a point $x$ in the domain of $f$ generates a small Macbeath region, then the corresponding point $p$ in the domain of $f^*$ generates a large Macbeath region. This allows us to employ a two-pronged sampling strategy, where samples are chosen either from $f$ or $f^*$, depending on the volumes of the Macbeath regions involved. We prove that this sampling strategy achieves the desired area-sensitive bound on the size of the hitting set used to form the approximation.

\bigskip

The remainder of the paper is organized as follows. In Section~\ref{sec:func-approx} we introduce our functional approach to approximation, where in Lemma~\ref{lem:outer} we prove that independent approximations to these functions can be combined to obtain a single approximation for the original convex body. We also present Lemma~\ref{lem:dual-outer}, which establishes how a hitting set for dual caps yields an approximation. Next, in Section~\ref{sec:cap-dual-cap} we explore the dual relationships between caps and dual caps. There we present Lemma~\ref{lem:dual-bases}, which shows that the projected bases of caps and dual caps are polars of each other. In Section~\ref{sec:hit-approx} we present our Macbeath-based sampling process. Finally, in Section~\ref{sec:additional} we present additional results, including a derivation of a similar area-sensitive approximation bound in the nonuniform setting and a proof of a technical lemma, which is used in Section~\ref{sec:macbeath}.

%=======================================================================
\section{Approximation Through a Functional Lens} \label{sec:func-approx}
%=======================================================================

%=======================================================================
\subsection{From Convex Bodies to Functions} \label{sec:support-sets}
%=======================================================================

In our analysis, we will use approximations of convex functions as intermediaries when working with approximations of convex bodies. In this section, we will describe this functional representation and its relevance to convex approximation. 

\begin{figure}[htbp]
    \centerline{\includegraphics[scale=0.4,page=2]{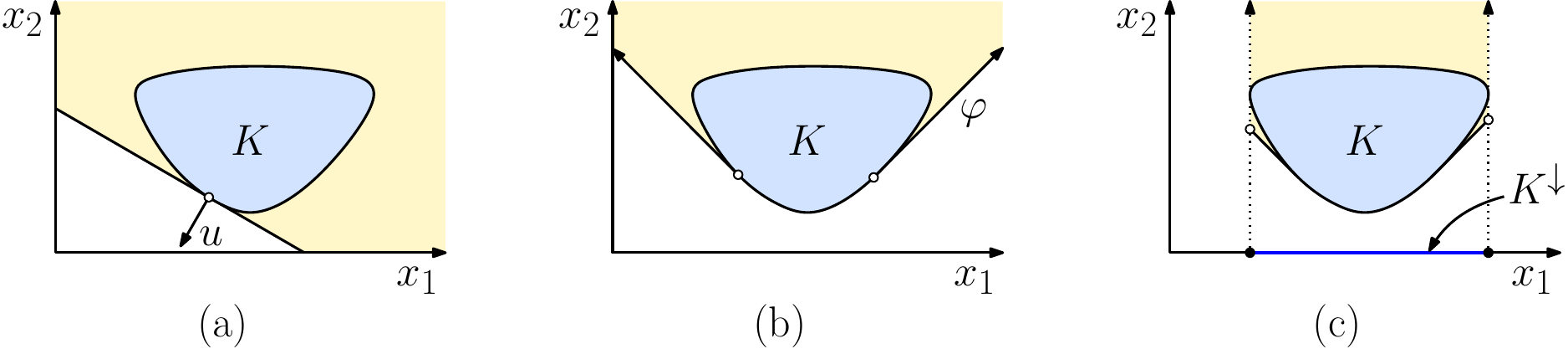}}
    \caption{A convex body $K$ and the associated convex function $\varphi$.}
    \label{fig:support-1-2}
\end{figure}

Given a convex body $K \subseteq \RE^d$, consider the convex set defined as the intersection of halfspaces $H$ with outer normal vector $u$ satisfying the following properties (see Figure~\ref{fig:support-1-2}(a)):
\begin{itemize}\setlength{\itemsep}{-0.5ex}\setlength{\parsep}{0pt}
\item $H$ contains $K$,
\item the last coordinate of $u$ is negative, and
\item the $L_{\infty}$ norm of $u$ is attained on its last coordinate.
\end{itemize}

%-----------------------------------------------------------------------
%-----------------------------------------------------------------------

Since this set is the intersection of the epigraphs of affine functions, it is the epigraph of a convex function $\varphi: \RE^{d-1} \to \RE$ (see Figure~\ref{fig:support-1-2}(b)). By construction, for any subgradient $p$ of this function, $\|p\| \leq \sqrt{d-1}$. Hence, by Lemma~\ref{lem:lipschitz} this function is Lipschitz continuous on the whole space $\RE^{d-1}$ with constant $\sqrt{d-1}$.

The function $\varphi$ has an unbounded domain, but for the sake of deriving area-sensitive bounds, it will be necessary to restrict its domain to one whose size more closely matches the surface of $K$. Given a set $K$ in $\RE^d$, define $K^{\downarrow}$ to be its orthogonal projection onto $\RE^{d-1}$. Since projection can only decrease areas, $\vol_{d-1}(K^{\downarrow})$ is not greater than $K$'s surface area, and so restricting the function to this domain would provide the desired area restriction (see Figure~\ref{fig:support-1-2}(c)). 

\begin{figure}[htbp]
    \centerline{\includegraphics[scale=0.4,page=1]{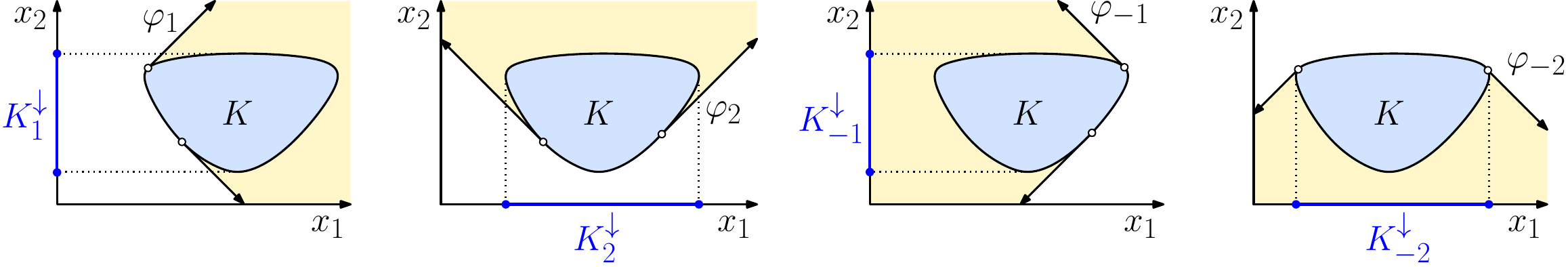}}
    \caption{The $2 d$ functions $\varphi_i$ defined by a convex body $K$.}
    \label{fig:support-1-1}
\end{figure}

In order to approximate $K$, we will apply this construction in multiple directions. Consider the set of $2 d$ vectors consisting of the coordinate unit vectors $e_i$ and their negations $-e_i$, for $1 \leq i \leq d$. We will define $2 d$ functions, where each of these vectors takes turns playing the role of the ``upward vertical'' axis, which reflects the value of the function. Let $\II$ denote the $(2 d)$-element index set $\{\pm 1, \ldots, \pm d\}$. For $i \in \II$, define $\varphi_i$ and $K^{\downarrow}_i$ as above (that is, as $\varphi$ and $K^{\downarrow}$, respectively), but when using terms like ``epigraph'' and ``projection'' the upward-directed vertical axis that records the function's value will be taken to be the $i$th coordinate vector, or its negation if $i < 0$ (see Figure~\ref{fig:support-1-1}).

%-----------------------------------------------------------------------
%-----------------------------------------------------------------------

It is not hard to see that, since every supporting hyperplane of $K$ will be included in one of these functions, the intersection of their epigraphs will equal $K$. The following lemma shows that this applies as well to outer $\eps$-approximations. However, it will be necessary to extend the domain of each function to a body that is slightly larger than the projection of $K$.

%-----------------------------------------------------------------------
\begin{lem} \label{lem:outer}
Let $K$ be a convex body in $\RE^d$ and let $\eps > 0$. For each $i \in \II$, let $\widehat{\varphi}_i$ be a convex lower $\eps$-approximation to $\varphi_i$ on $K^{\downarrow}_i \oplus \eps B^{d-1}_2$. Then $\bigcap_{i \in \II} (\epi \widehat{\varphi}_i)$ is an outer convex $\eps$-approximation to $K$.
\end{lem}
%-----------------------------------------------------------------------

%-----------------------------------------------------------------------
\begin{proof}
Let $P$ denote the intersection of the epigraphs of $\widehat{\varphi}_i$, for $i \in \II$. For any $i \in \II$, since $\widehat{\varphi}_i$ is a convex lower approximation to $\varphi_i$, we have
\[
    \epi \widehat{\varphi}_i
        ~ \supseteq ~ \epi \varphi_i
        ~ \supseteq ~ K, ~~ \text{for all $i \in \II$,}
\]
which implies that $P \supseteq K$, and hence $P$ is an outer approximation to $K$.

To complete the proof, we show that $P$ is an $\eps$-approximation to $K$ by showing that any point $u$ that lies at distance greater than $\eps$ from $K$ does not lie within $P$. Let $v$ be the closest point of $K$ to $u$ (see Figure~\ref{fig:outer}). Let $i$ denote the index of the coordinate of $u - v$ where the $L_{\infty}$ norm is attained, and negate $i$ if the value of this coordinate is positive. Thus, $i \in \II$. For the remainder of the proof, we take $i$ to be the last coordinate (negated if $i < 0$). Let $u = (x; s)$ and $v = (y; t)$. Let $u' = (x'; s')$ be the point on the segment $\overline{u v}$, such that $\|u' - v\| = \eps$. 

%-----------------------------------------------------------------------
\begin{figure}[htbp]
    \centerline{\includegraphics[scale=0.4]{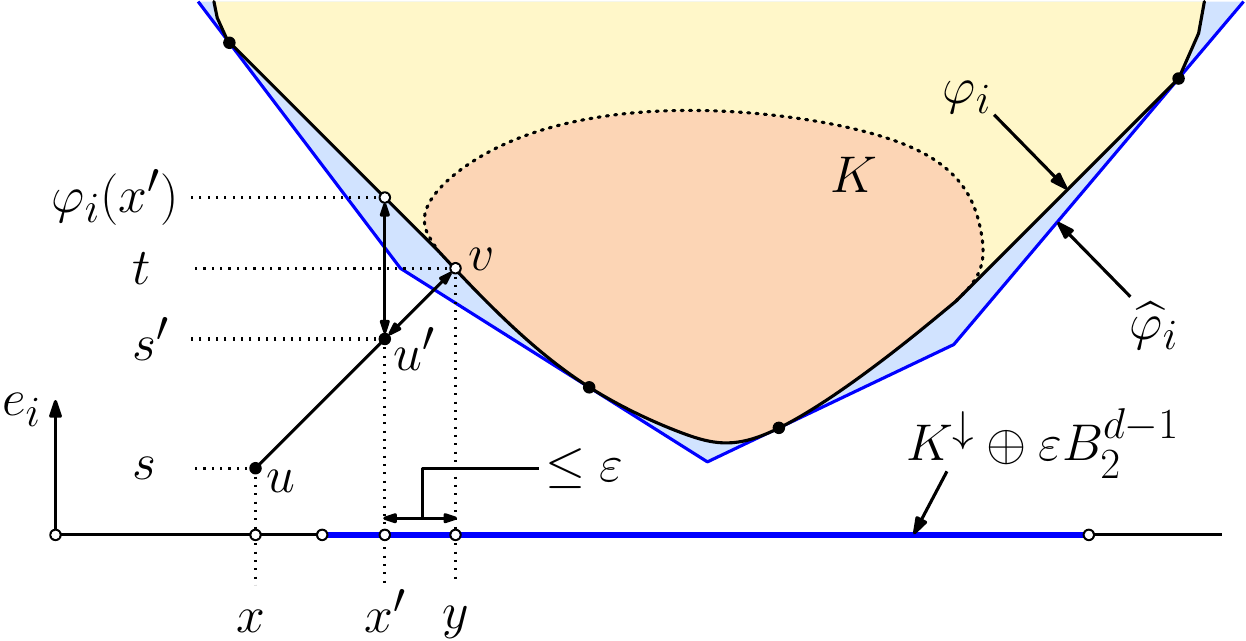}}
    \caption{Proof of Lemma~\ref{lem:outer}.}
    \label{fig:outer}
\end{figure}
%-----------------------------------------------------------------------

Clearly, $|x' - y| \leq \eps$. Since $y \in K^{\downarrow}_i$, we have $x' \in K^{\downarrow}_i \oplus \eps B^{d-1}_2$. By local-minimality considerations, $v$ is the closest point to $u'$ on $K^+_i$, and hence $\varphi_i(x') - s' \geq \eps$. Since $\widehat{\varphi}_i$ is a convex lower $\eps$-approximation to $\varphi_i$, we have
\[
    \widehat{\varphi}_i(x')
        ~ \geq ~ \varphi_i(x') - \eps
        ~ \geq ~ s'.
\]
In summary, $v \in \epi \varphi_i$ and $u' \notin \epi \widehat{\varphi}_i$, and by the convexity of $\epi \widehat{\varphi}_i$, it follows that $u \notin \epi \widehat{\varphi}_i$. This implies that $u$ is external to $P$, as desired.
\end{proof}
%-----------------------------------------------------------------------

Later in the paper, we will prove Theorem~\ref{thm:func-approx}. Combining that with the above lemma nearly establishes Theorem~\ref{thm:main}. To combine the two, we need to relate $\arad(K)$ with $\vrad(D)$, where $D = K^{\downarrow}_i \oplus \eps B^{d-1}_2$. The following lemma will be useful in establishing this relation. It shows that expanding such a body by distance $\eps$ does not increase the volume by more than a constant factor.

%-----------------------------------------------------------------------
\begin{lem} \label{lem:expansion}
Let $K$ be a convex body in $\RE^d$ of minimum width at least $\eps$. Then
\[
    \vol_d\big( K \oplus \eps B^d_2 \big)
        ~ \leq ~ c_d \cdot \vol_d(K),
\]
where $c_d$ is a constant depending on $d$.
\end{lem}
%-----------------------------------------------------------------------

%-----------------------------------------------------------------------
\begin{proof}
By Steinhagen’s lemma~\cite{BeH92}, a convex body in $\RE^d$ of minimum width $w$ contains a ball of radius at least $w/(4 \sqrt{d})$. This implies that $K$ contains a ball of radius $\eps/(4 \sqrt{d})$ centered at some point $x \in K$. Thus, a uniform scaling of $K$ about $x$ by a factor of
\[
    \frac{\eps + \eps/(4\sqrt{d})}{\eps/(4\sqrt{d})}
        ~ = ~ 1 + 4 \sqrt{d}
\]
contains $K \oplus \eps B^d_2$. We have $\vol_{d}\big( K \oplus \eps B^d_2 \big) \leq \big( 1 + 4 \sqrt{d} \big)^d \cdot \vol_d(K)$, as desired.
\end{proof}
%-----------------------------------------------------------------------

Since $K$ has minimum width at least $\eps$, this applies to $K^{\downarrow}_i$ as well, for any $i \in \II$. Thus, by the above and the fact that orthogonal projection cannot increase areas, we have
\[
    \vol_{d-1}\big( K^{\downarrow}_i \oplus \eps B^{d-1}_2 \big) 
        ~ \leq ~ c_{d-1} \cdot \vol_{d-1}(K^{\downarrow}_i)
        ~ \leq ~ c_{d-1} \cdot \area(K).
\]
Defining the domain $D_i = K^{\downarrow}_i \oplus \eps B^{d-1}_2$, we have
\[
    \vrad(D_i)
        ~ =    ~ \left( \frac{\vol_{d-1}(D_i)}{\vol_{d-1}\big( B^{d-1}_2 \big)} \right)^{\kern-2pt \frac{1}{d-1}}
        ~ \leq ~ \left( \frac{c_{d-1} \cdot \area(K)}{\vol_{d-1}\big( B^{d-1}_2 \big)} \right)^{\kern-2pt \frac{1}{d-1}}.
\]
By definition,
\[
    \arad(K)
        ~ = ~ \left( \frac{\area(K)}{\area(B^d_2)} \right)^{\kern-2pt \frac{1}{d-1}},
\]
and hence
\[
    \vrad(D_i)
        ~ \leq ~ c'_d \cdot \arad(K), 
        \quad\text{where~~} c'_d ~ = ~ \left(\frac{c_{d-1} \cdot \area\big( B^d_2 \big)}{\vol_{d-1}\big( B^{d-1}_2 \big)}\right)^{\kern-2pt \frac{1}{d-1}}.
\]

Therefore, to establish Theorem~\ref{thm:main}, we can apply Theorem~\ref{thm:func-approx} to approximate each of the functions $\varphi_i$ on the expanded domains $D_i$, for each $i \in \II$ (exploiting the fact that they all have bounded Lipschitz constants), and then we invoke Lemma~\ref{lem:outer} to combine these $2 d$ function approximations to approximate $K$. The remainder of the paper will focus on proving Theorem~\ref{thm:func-approx}.

%=======================================================================
\subsection{Regularity Assumptions} \label{sec:regularity}
%=======================================================================

Our proof of Theorem~\ref{thm:func-approx} will make use of the Legendre transform. Consider a lower semicontinuous convex function $f$ on $\RE^{d-1}$, and let $U$ be an open convex set. Following Rockafellar~\cite[Section 26]{Roc97}, $(U,f)$ is said to be of \emph{Legendre type} if $U$ is nonempty, $f$ is differentiable and strictly convex throughout $U$, and whenever $x_1, x_2, \ldots$ is a sequence in $U$ converging to a point $x$ on the boundary of $U$,
\[
    \lim_{i \rightarrow \infty} \| \Gradient f(x_i) \| 
        ~ = ~ +\infty.
\]
Throughout, we will work with such functions with the additional constraint that they are Lipschitz continuous over most of their domain. This is encapsulated in the following concept.

\begin{definition}[Regularity Assumptions] \label{def:reg-assump}
Given an extended real-valued function $f$ on $\RE^{d-1}$, a convex body $D$ in $\RE^{d-1}$, and positive parameters $\eps$ and $\lambda$, $f$ satisfies the \emph{Regularity Assumptions} with respect to $D$, $\eps$, and $\lambda$ if the following hold:
\begin{enumerate}
\item[$(i)$] It is lower semicontinuous, convex, and letting $U$ denote $\interior(\dom f)$, $U$ is bounded and $(U,f)$ is of Legendre type,

\item[$(ii)$] The origin lies within the interior of $D$ and $D \oplus \eps B^{d-1}_2 \subseteq U \subseteq D \oplus 2\eps B^{d-1}_2$, and

\item[$(iii)$] $f$ is Lipschitz continuous on $D \oplus \eps B^{d-1}_2$, with Lipschitz constant $\lambda$.
\end{enumerate}
\end{definition}

In the remainder of the paper, we establish the following area-sensitive approximation for such functions.

%-----------------------------------------------------------------------
\begin{thm}[Regularized Functional Approximation] \label{thm:reg-func-approx} 
Let $D$ be a compact convex domain in $\RE^{d-1}$ of minimal width at least $\eps$, and let $f$ be a lower semicontinuous convex function on $\RE^{d-1}$ satisfying the Regularity Assumptions with respect to $D$ and positive parameters $\eps$ and $\lambda$. Then, for some constant $c_d$ (depending on $d$) there exists a set of at most
\[
    c_d \left( \max(1, \lambda) \cdot \frac{\vrad(D)}{\eps} \right)^{\kern-2pt \frac{d-1}{2}}
\]
affine functions on $\RE^{d-1}$ such that the pointwise maximum of these functions is a convex lower $\eps$-approximation to $f$ on $D$.
\end{thm}
%-----------------------------------------------------------------------

%-----------------------------------------------------------------------
\begin{lem} \label{lem:regularizing}
Theorem~\ref{thm:func-approx} follows from Theorem~\ref{thm:reg-func-approx} (subject to an adjustment of the constant $c_d$).
\end{lem}
%-----------------------------------------------------------------------

%-----------------------------------------------------------------------
\begin{proof}
Let $D$, $f$, $\eps$, and $\lambda$, be as specified in Theorem~\ref{thm:func-approx}. Let $\eps' = \eps/2$, and let $\delta = \eps/(4+2\lambda)$. We will show how to perturb $f$ to a function $\dbtilde{f}$ that satisfies the \hyperref[def:reg-assump]{Regularity Assumptions} with the Lipschitz constant $\lambda' = \lambda+1$, so that any lower $\eps'$-approximation to $\dbtilde{f}$ on $D$ is a lower $\eps$-approximation to $f$ on $D$.

First, apply a translation such that the origin lies in the interior of $D$. Let $D' = D \oplus \frac{3}{2} \eps B^{d-1}_2$. By applying Kirszbraun's theorem~\cite{Fed96}, we can extend the domain of $f$ to $D'$ while maintaining the same Lipschitz constant $\lambda$ (see Figure~\ref{fig:support-2}(a)).

%-----------------------------------------------------------------------
\begin{figure}[htbp]
    \centerline{\includegraphics[scale=0.4,page=1]{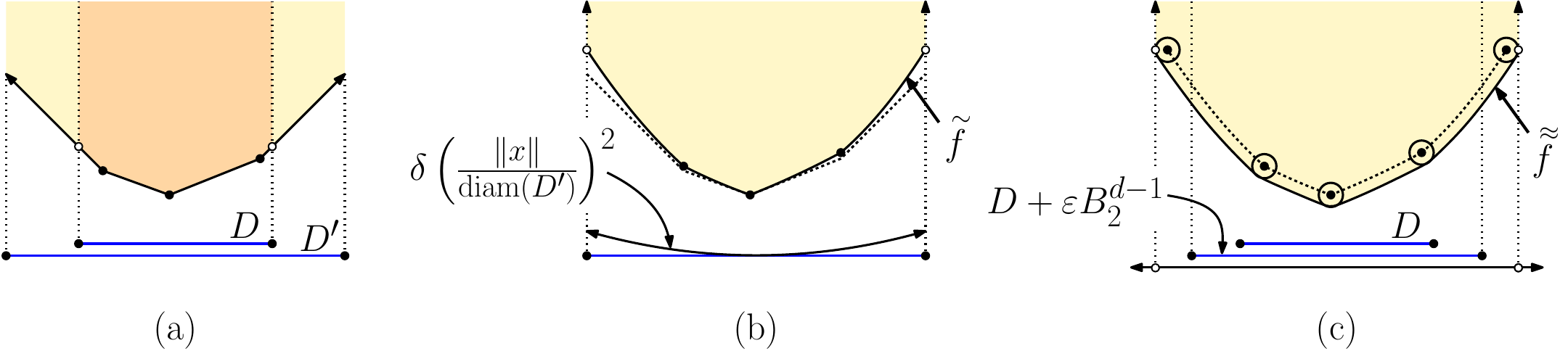}}
    \caption{Proof of Lemma~\ref{lem:regularizing}.}
    \label{fig:support-2}
\end{figure}
%-----------------------------------------------------------------------

To make $f$ strictly convex, we take its sum with a slowly growing quadratic function. For $x \in D'$, define
\[
    \simtilde{f}(x)
        ~ = ~ f(x) + \delta \left(\frac{\|x\|}{\diam(D')}\right)^{\kern-2pt 2}
\]
(see Figure~\ref{fig:support-2}(b)). Observe that this has the effect of increasing the function values throughout $D'$ by at most $\delta$. The Lipschitz constant of the quadratic term is at most 
\[
    \frac{2 \delta}{\diam(D')^2} \cdot \sup_{x \in D'} \|x\| 
        ~ \leq ~ \frac{\eps}{\diam(D')}
        ~ <    ~ 1,
\]
where we have used the facts that $\delta \leq \eps/2$ and $\diam(D') > \eps$. Thus, $\simtilde{f}$ is Lipschitz continuous on $D'$ with Lipschitz constant $\lambda + 1 = \lambda'$.

Next, to smooth this function while keeping its support bounded, we convolve it with a smooth function of bounded support. For any $v \in \delta B^{d-1}_2$, define $b_{\delta}(v) = -\sqrt{\delta^2 - \|v\|^2}$. The graph of this function is the lower hemisphere of the ball $\delta B^d_2$. For any $x \in D' \oplus \delta B^{d-1}_2$, define
\[
    \dbtilde{f}(x) 
        ~ = ~ \inf_{y \in D' \cap \big( x + \delta B^{d-1}_2 \big)} \left( \simtilde{f}(y) + b_{\delta}(x-y) \right)
\]
(see Figure~\ref{fig:support-2}(c)). This function is the infimal convolution of $\simtilde{f}$ with $b_\delta$. We can relate its epigraph to the epigraph of $\simtilde{f}$ as follows. The epigraph of the Kirszbraun extension of $f$ is a convex set whose boundary extends upward to $\infty$ above the points of $\bd D'$. This also applies to $\simtilde{f}$. Convolving $\simtilde{f}$ with $b_{\delta}$ effectively grows the epigraph by taking its Minkowski sum with the Euclidean ball of radius $\delta$. Clearly, $\dbtilde{f}$ is a smooth and strictly convex function, and $\dom \dbtilde{f} = D' \oplus \delta B^{d-1}_2$. Since $\delta \leq \eps/2$, this is contained within $D \oplus 2 \eps B^{d-1}_2$, and thus, \hyperref[def:reg-assump]{Regularity Assumption}~(ii) is satisfied. 

We claim that $\dbtilde{f}$ is Lipschitz continuous with constant $\lambda'$ on $D \oplus \eps B^{d-1}_2$. To see this, consider any $x \in D \oplus \eps B^{d-1}_2$, and let $h$ denote the associated supporting hyperplane at the point $(x; \dbtilde{f}(x))$ on the epigraph of $\dbtilde{f}$. Since this epigraph is the Minkowski sum of the epigraph of $\simtilde{f}$ with the Euclidean ball of radius $\delta$, each of its supporting hyperplanes is the translate of a supporting hyperplane of $\simtilde{f}$ by distance $\delta$ along the hyperplane's outward normal vector. Since $x$ is at distance at least $\delta$ from $\bd D'$, it follows that $h$ arises from a point in the epigraph of $\simtilde{f}$ that lies within $D'$. Therefore, by the Lipschitz continuity of $\simtilde{f}$ on $D'$, the same Lipschitz condition holds for $\dbtilde{f}$ on $D \oplus \eps B^{d-1}_2$. This establishes \hyperref[def:reg-assump]{Regularity Assumption}~(iii).

To show that $\dbtilde{f}$ is of Legendre type, let $U = \interior(\dom f)$. Consider any point $x \in \bd U$, and let $x_1, x_2, \ldots$ be a sequence in $U$ converging to $x$. Let $y$ be the closest point of $D'$ to $x$. Clearly, $\|x - y\| = \delta$, and by the convexity of $D'$, $y$ is the only point of $D' \cap \big( x + \delta B^{d-1}_2 \big)$ that contributes to the term $b_{\delta}(x-y)$ in the definition of $\dbtilde{f}(x)$. For the sake of computing the limit in the definition of Legendre type, we may assume that the points of the sequence $x_1, x_2, \ldots$ lie entirely within $x + \delta B^{d-1}_2$. It is easy to verify that for any $v \in \delta B^{d-1}_2$,
\[
    \|\Gradient b_{\delta}(v)\|
        ~ = ~ \frac{\|v\|}{\sqrt{\delta^2 - \|v\|^2}},
\]
which tends to $+\infty$ as $\|v\| \rightarrow \delta$. As $i \rightarrow \infty$, $\|x_i - y\|$ approaches $\delta$, and hence $\|\Gradient b_{\delta}(x_i - y)\| \rightarrow +\infty$. Since $\Gradient \simtilde{f}(x)$ is bounded throughout $D'$, as we approach the boundary of $U$, $\big\|\Gradient \dbtilde{f} \SP \big\|$ is dominated by $\|\Gradient b_{\delta}\|$. Thus, $(U,\dbtilde{f})$ is of Legendre type, satisfying \hyperref[def:reg-assump]{Regularity Assumption}~(i).

In order to obtain the desired approximation, we invoke Theorem~\ref{thm:reg-func-approx} on our regularized function $\dbtilde{f}$ with parameters $\lambda' = \lambda+1$ and $\eps' = \eps/2$. Since $\min(1, \lambda') \leq 2 \cdot \min(1, \lambda)$, the increase in the complexity bound is at most $(2 \cdot 2)^{(d-1)/2} = 2^{(d-1)}$, which can be absorbed by an adjustment in the constant $c_d$. Let $\widehat{f}$ be the resulting lower approximation to $\dbtilde{f}$. We assert that this is a lower $\eps$-approximation to the original function $f$. To see this, observe that since $b_{\delta}$ is nonpositive throughout its domain, the convolution with $b_{\delta}$ cannot increase function values. Since $\simtilde{f}$ has Lipschitz constant $\lambda' = \lambda+1$, it follows from basic geometry and our choice of $\delta$ that the decrease in the value of the function at any point is at most 
\[
    \delta \sqrt{1 + (\lambda+1)^2} 
        ~ \leq ~ \delta (2 + \lambda)
        ~ \leq ~ \frac{\eps}{2}.
\]
Also, observe that the amount of decrease is at least $\delta$ (which arises when the gradient is zero). Combining this with the positive error of at most $\delta$ induced with $\simtilde{f}$, for all $x \in D$ we have 
\[
    0
        ~ \leq ~ f(x) - \dbtilde{f}(x)
        ~ \leq ~ \frac{\eps}{2}.
\]
Thus, for all $x \in D$,
\[
    f(x) - \widehat{f}(x)
        ~ =    ~ \left( f(x) - \dbtilde{f}(x) \right) + \left( \dbtilde{f}(x) - \widehat{f}(x) \right)
        ~ \in ~ \left[ 0 + 0, \frac{\eps}{2} + \eps' \right]
        ~ =   ~ [0, \eps],
\]
which implies that $\widehat{f}$ is a lower $\eps$-approximation to $f$, as desired.
\end{proof}
%-----------------------------------------------------------------------

%=======================================================================
\section{Dual Caps and Caps} \label{sec:cap-dual-cap}
%=======================================================================

%=======================================================================
\subsection{Basic Concepts} \label{sec:cap-basics}
%=======================================================================

In this section, we introduce the notion of caps and dual caps in our functional context and explore their properties. We begin by defining the concept of a dual cap. Given a closed convex set $K$ and a point $a \notin K$, define the \emph{dual cap} as the set of boundary points $x \in \bd K$ such that the interior of some supporting halfspace of $K$ at $x$ does not contain $a$ (recall Figure~\ref{fig:methods}). Equivalently, this is the set of points on the boundary of $K$ that are ``visible'' from $a$ in the sense that they lie within the smallest cone with apex $a$ that contains $K$, $\Cone(K, a) = \{ a + \gamma (z-a) \ST z \in K ~\text{and}~ \gamma \geq 0 \}$.

\begin{figure}[htbp]
    \centerline{\includegraphics[scale=0.4,page=1]{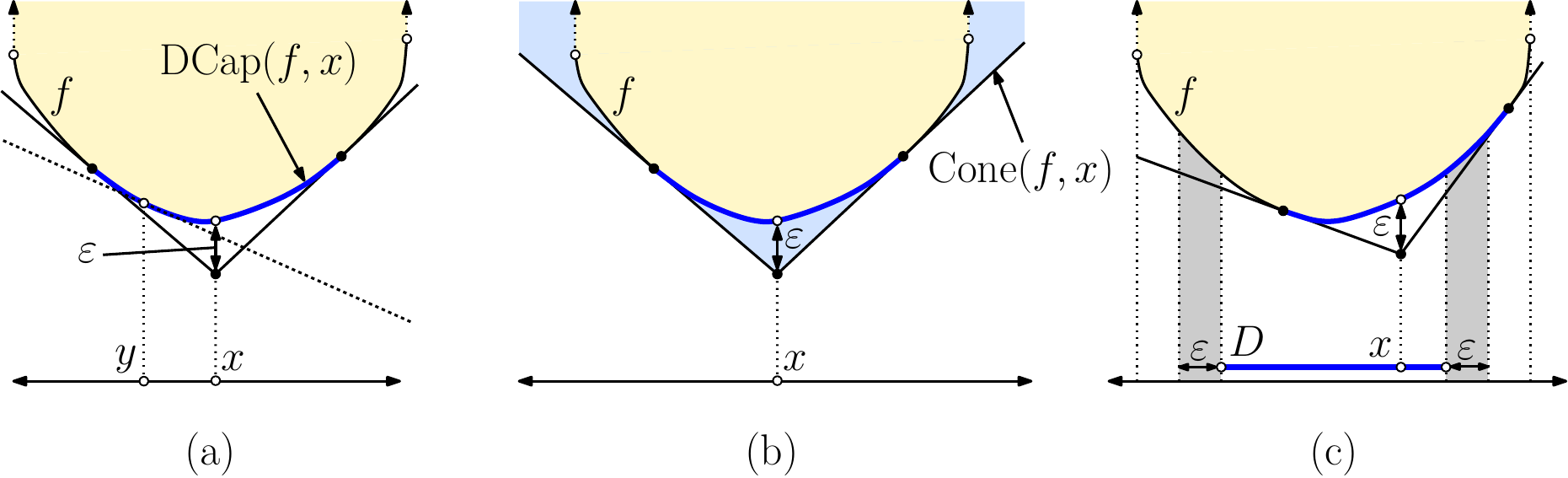}}
    \caption{(a) Dual cap $\DCap(f, x)$, (b) $\Cone(f, x)$, and (c) a useful dual cap.}
    \label{fig:cap-dual-cap-1}
\end{figure}

We can easily adapt this concept to our functional setting. Given a lower semicontinuous convex function $f$ on $\RE^{d-1}$ and $x \in \interior(\dom f)$, we define the \emph{dual cap} of $f$ at $x$ to be the dual cap of the convex set $\epi f$ at the point $(x; f(x) - \eps)$. That is, it consists of points on $\graph f$ such that there exists a supporting hyperplane at this point that passes on or above $(x; f(x) - \eps)$:
\[
    \DCap(f, x)
        ~ = ~ \big\{ (y; f(y)) \ST \exists p \in \partial f(y), f(y) + \inner{p}{x - y} \geq f(x) - \eps \big\}
\]
(see Figure~\ref{fig:cap-dual-cap-1}(a)). 

%-----------------------------------------------------------------------
%-----------------------------------------------------------------------

Again, we can relate the dual cap of a convex function $f$ to a cone of visible points on $f$'s graph. Define $\Cone(f, x) = \Cone(\epi f , (x; f(x) - \eps))$ (see Figure~\ref{fig:cap-dual-cap-1}(b)). It is easy to see that every supporting hyperplane of $\epi f$ whose upper halfspace does not contain $(x; f(x) - \eps)$ intersects every ray of $\Cone(f, x)$. We have the following direct consequence.

%-----------------------------------------------------------------------
\begin{lem} \label{lem:dbase-cone}
Let $f$ be a lower semicontinuous convex function on $\RE^{d-1}$, and let $x \in \interior(\dom f)$. Then
\[
    \Cone(f, x)
        ~ = ~ \Cone(\DCap(f, x), (x; f(x)-\eps)).
\]
\end{lem}
%-----------------------------------------------------------------------

Difficulties arise in the analysis of dual caps whose defining point $x$ is very close to the boundary of $\dom f$. To address this, we focus on functions that satisfy \hyperref[def:reg-assump]{Regularity Assumptions}. Recalling that for such functions, $\dom f$ contains the $\eps$-expansion of a convex body $D$, we restrict our attention to the dual caps induced by points in $D$.

%-----------------------------------------------------------------------
\begin{definition}[Useful Point/Dual Cap] \label{def:useful}
Given a function $f$ that satisfies the \hyperref[def:reg-assump]{Regularity Assumptions} with respect to a convex domain $D$, we say that a point $x \in \dom f$ is \emph{useful} if $x \in D$, and we say that a dual cap, $\DCap(f,x)$, is \emph{useful} if $x$ is useful (see Figure~\ref{fig:cap-dual-cap-1}(c)).
\end{definition}
%-----------------------------------------------------------------------

The relevance of useful points to function approximation can be understood through the concept of hitting sets. Given a function $f$ that satisfies the \hyperref[def:reg-assump]{Regularity Assumptions}, we say that a discrete set $\mathcal{H} \subset \interior(\dom f)$ is a \emph{hitting set} for the set of useful dual caps if $\DCap(f, x)^{\downarrow} \cap \mathcal{H} \neq \emptyset$, whenever $x$ is useful. This implies that there exists $y \in \mathcal{H}$ such that the supporting hyperplane to $\epi f$ at $(y; f(y))$ separates $(x; f(x))$ from $(x; f(x) - \eps)$. Thus, we have the following.

%-----------------------------------------------------------------------
\begin{lem} \label{lem:dual-outer}
Given a function $f$ that satisfies the \hyperref[def:reg-assump]{Regularity Assumptions} with respect to a convex domain $D$ and $\eps > 0$, let $\mathcal{H} \subset \interior(\dom f)$ be a hitting set for the set of useful dual caps. For each $y \in \mathcal{H}$, let $H^+(y)$ denote the closed upper halfspace defined by the supporting hyperplane to $\epi f$ at $(y; f(y))$. Then $\bigcap_{y \in \mathcal{H}} H^+(y)$ is a piecewise-linear convex lower $\eps$-approximation to $f$ on $D$.
\end{lem}
%-----------------------------------------------------------------------

Our construction of these hitting sets, which will be described in Section~\ref{sec:hit-approx}, is based on sampling points from certain regions (specifically Macbeath regions) that lie near the lower boundary of the epigraph of the function and the function's dual (defined in Section~\ref{sec:duality}). Our analysis of these hitting sets will involve the concept of a cap of a function, which we define next.

\begin{figure}[htbp]
    \centerline{\includegraphics[scale=0.4,page=2]{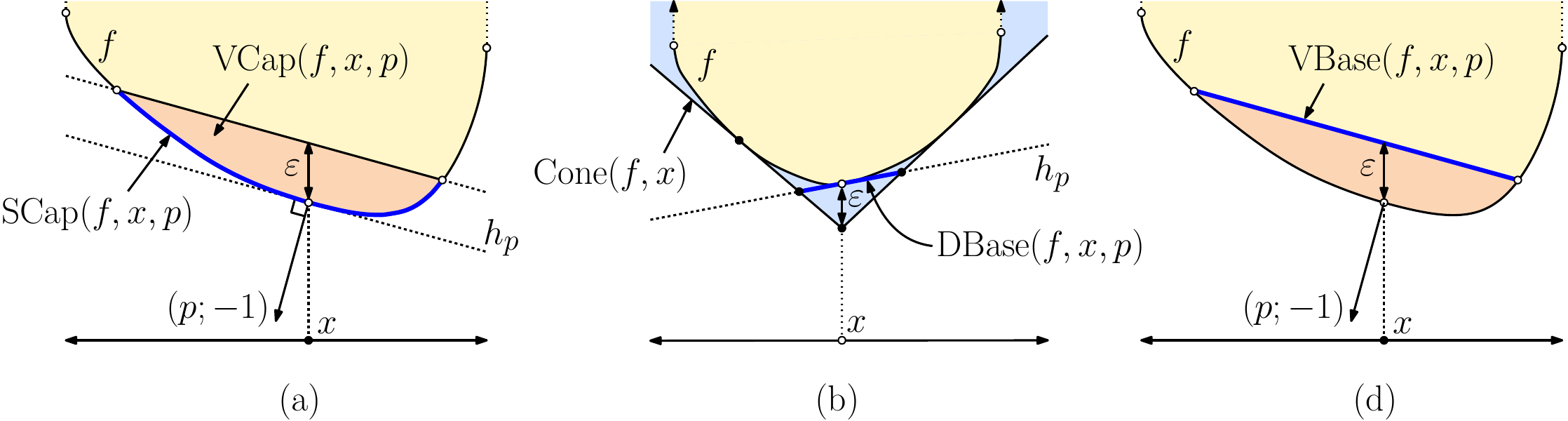}}
    \caption{(a) Volume and surface caps, (b) the base of a dual cap, and (c) the base of a cap.}
    \label{fig:cap-dual-cap-2}
\end{figure}

Given a closed convex set $K$ and any halfspace $H$ that intersects $K$, we define a \emph{volume cap} to be $K \cap H$ and a \emph{surface cap} to be $H$'s intersection with $K$'s boundary. To adapt these concepts to our functional setting, consider a lower semicontinuous convex function $f$ on $\RE^{d-1}$ and $x \in \interior(\dom f)$. Recall that $\subgrad f(x)$ denotes the set of all subgradients at $x$. Given $p \in \subgrad f(x)$, let $h_p$ denote the hyperplane supporting $\epi f$ at $(x; f(x))$ with outer normal $(p; -1)$, which exists by Lemma~\ref{lem:subgradient}. Let $H_p$ be the closed lower halfspace bounded by $h_p$. Letting $e_d$ denote the $d$th unit vector, define the \emph{volume cap} and \emph{surface cap} of $f$ for $x$ and $p$ to be
\[
    \VCap(f, x, p)
        ~ = ~ \epi f \cap (H_p + \eps \SP e_d)
    \qquad\text{and}\qquad
    \SCap(f, x, p)
        ~ = ~ \graph f \cap (H_p + \eps \SP e_d),
\]
respectively (see Figure~\ref{fig:cap-dual-cap-2}(a)). 

%-----------------------------------------------------------------------
%-----------------------------------------------------------------------

It will also be useful to introduce two related flat structures. Define the associated \emph{dual-cap base} and \emph{volume-cap base} as 
\[
    \DBase(f, x, p) 
        ~ = ~ h_p \cap \Cone(f, x)
    \qquad\text{and}\qquad
    \VBase(f, x, p) 
        ~ = ~ (h_p + \eps \SP e_d) \cap \epi f
\]
(see Figures~\ref{fig:cap-dual-cap-2}(b) and (c)).

There are connections between these various cap structures. Given $f$, $x$, and $p$ as above, it is known from standard results in convexity theory that $\VCap(f, x, p)$ contains an ellipsoid whose volume is within a constant factor of the cap's volume~\cite{Bal91, Joh48}. In the next lemma, we show that there exists such an ellipsoid with the additional property that its vertical projection is contained within the vertical projection of $\DBase(f, x, p)$ and their projected volumes are comparable (see Figure~\ref{fig:cap-ratio-lemma}(a)).

%-----------------------------------------------------------------------
\begin{lem} \label{lem:cap-ratio}
Let $f$ be a lower semicontinuous convex function on $\RE^{d-1}$ with a bounded effective domain, let $x \in \interior(\dom f)$, and let $p \in \subgrad f(x)$. There exists an ellipsoid $E \subseteq \VCap(f, x, p)$ such that $E^{\downarrow} \subseteq \DBase(f, x, p)^{\downarrow}$ and
\[
    \vol_d(E) 
        ~ \geq ~ c \cdot \vol_d(\VCap(f, x, p)) 
        ~ \geq ~ c' \eps \cdot \vol_{d-1}(\DBase(f, x, p)^{\downarrow}),
\]
where $c$ and $c'$ are constants depending on the dimension.
\end{lem}
%-----------------------------------------------------------------------

%-----------------------------------------------------------------------
\begin{proof}
Recall from Lemma~\ref{lem:dbase-cone} that $\Cone(f, x) = \Cone(\DCap(f, x), (x; f(x)-\eps))$. Let $h_p$ denote the hyperplane supporting $\epi f$ at $(x; f(x))$ with outer normal $(p; -1)$, and let $H_p$ denote the closed lower halfspace bounded by $h_p$. Let $T$ denote the portion of this cone that lies on or below $h_p$, that is, $T = H_p \cap \Cone(f,x)$. Let $T'$ be a scaling of $T$ by a factor of $2$ about $(x; f(x) - \eps)$ (see Figure~\ref{fig:cap-ratio-lemma}(b)). We first show the following straightforward facts about this construction:
\begin{enumerate}\setlength{\itemsep}{-0.5ex}\setlength{\parsep}{0pt}
\item[$(i)$] $T + \eps \SP e_d ~\subseteq~ \VCap(f, x, p) ~\subseteq~ T'$.
\item[$(ii)$] $\vol_d(T) ~ = ~ \big( \frac{\eps}{d} \cdot \vol_{d-1}(\DBase(f, x, p)^{\downarrow} \big)$.
\item[$(iii)$] $T^{\downarrow} = \DBase(f, x, p)^{\downarrow}$.
\end{enumerate}

%-----------------------------------------------------------------------
\begin{figure}[htbp]
    \centerline{\includegraphics[scale=0.4,page=1]{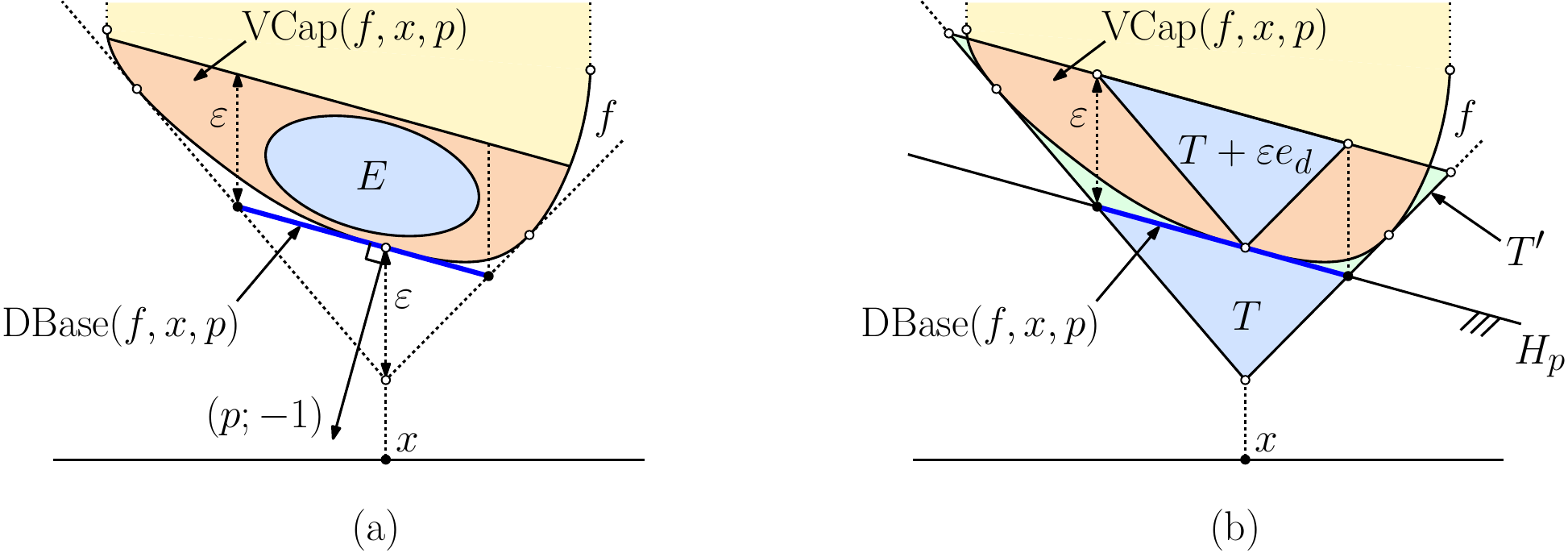}}
    \caption{Lemma~\ref{lem:cap-ratio}.}
    \label{fig:cap-ratio-lemma}
\end{figure}
%-----------------------------------------------------------------------

Assertion~(i) follows easily by convexity and the definition of $\VCap(f, x, p)$. To prove assertion~(ii), observe that $T$ is a bounded cone whose base is the intersection of $h_p$ with $\Cone(f, x)$, which is just $\DBase(f, x, p)$. The vertical distance between $T$'s apex, $(x; f(x) - \eps)$, and its base is $\eps$. Therefore, by basic geometry, its volume is
\[
    \vol_d(T)
        ~ = ~ \frac{\eps}{d} \vol_{d-1}\left( \DBase(f, x, p)^{\downarrow} \right).
\]
Assertion (iii) follows from the fact that $T$'s apex lies below its base and therefore $T$'s vertical projection is just the vertical projection of its base, $\DBase(f, x, p)^{\downarrow}$.

Returning to the proof, let $E$ be the maximum volume ellipsoid contained in $T + \eps \SP e_d$. From the first containment of~(i), we have 
\[
    E 
        ~ \subseteq ~ T + \eps \SP e_d 
        ~ \subseteq ~ \VCap(f, x, p).
\]
By Ball's bound on the volume ratio~\cite{Bal91}, which relates the volume of a convex body to that of its maximal volume ellipsoid, there exists an absolute constant $\alpha > 0$ independent of the dimension such that $\vol_d(E) \geq \vol_d(T)/\big( \alpha \sqrt{d} \big)^d$. From the second containment of~(i), we have $\VCap(f, x, p) \subseteq T'$. Since $T'$ is a factor-2 scaling of $T$, $\vol_d(T) = \vol_d(T')/2^d$. Therefore, by setting $c = 1/(2 \alpha \sqrt{d})^d$, we have 
\[
    \vol_d(E) 
        ~ \geq ~ \frac{\vol_d(T)}{(\alpha \sqrt{d})^d}
        ~ =    ~ \frac{\vol_d(T')}{(2 \alpha \sqrt{d})^d}
        ~ \geq ~ \frac{\vol_d(\VCap(f, x, p))}{(2 \alpha \sqrt{d})^d}
        ~ =    ~ c \cdot \vol_d(\VCap(f, x, p)).
\]
This establishes the first inequality in the statement of the lemma, and the second inequality follows from assertion~(ii) and setting $c' = c/d$. Combining the fact that $E \subseteq T + \eps \SP e_d$ and (iii) implies that $E^{\downarrow} \subseteq \DBase(f, x, p)^{\downarrow}$.
\end{proof}
%-----------------------------------------------------------------------

%=======================================================================
\subsection{Duality Transforms} \label{sec:duality}
%=======================================================================

Throughout the paper, we will make use of two well-known dual transforms, the polar transform of a convex body and the dual conjugate of a convex function. In this section, we review these concepts and their relevant properties.

\begin{figure}[htbp]
    \centerline{\includegraphics[scale=0.4]{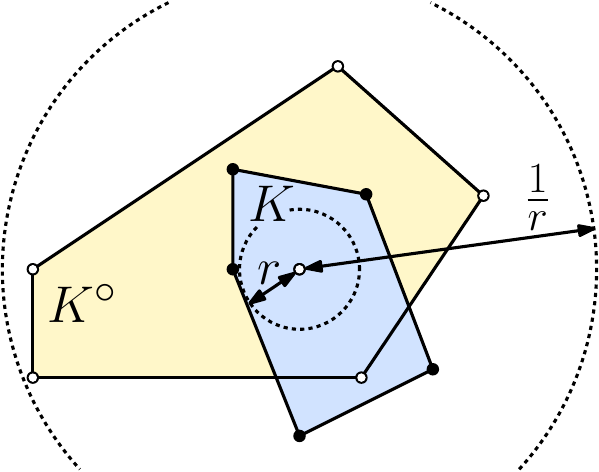}}
    \caption{The polar body of $K$.}
    \label{fig:polar}
\end{figure}

Given a convex body $K \subseteq \RE^d$ that contains the origin in its interior, its \emph{polar}, denoted here $\stdpolar{K}$, is defined to be $\{u \in \RE^d \ST \inner{u}{v} \leq 1, \forall v \in K\}$. Given $\alpha > 0$, let $\alpha \stdpolar{K}$ denote a scaling of $\stdpolar{K}$ by a factor of $\alpha$. Clearly, $\alpha \stdpolar{K} = \{u \in \RE^d \ST \inner{u}{v} \leq \alpha, \forall v \in K\}$. It can be easily verified that a body and its polar have a reciprocal relationship in the sense that if $K$ contains the Euclidean ball of radius $r$ centered at the origin, then $\stdpolar{K}$ is contained within the Euclidean ball of radius $1/r$ centered at the origin (see Figure~\ref{fig:polar}).

%-----------------------------------------------------------------------
%-----------------------------------------------------------------------

An important concept from the theory of convex sets, called the \emph{Mahler volume}, states that given a convex body $K$, the product of the volumes of $K$ and $\stdpolar{K}$ is bounded below by a constant depending only on the dimension (see, e.g., \cite{BoM87,Kup08,Naz12}). We state the following result in $\RE^{d-1}$, which is where we will apply it.

%-----------------------------------------------------------------------
\begin{lem}[Kuperberg~\cite{Kup08}] \label{lem:mahler-bounds}
Given a convex body $K \subseteq \RE^{d-1}$ whose interior contains the origin, $\vol_{d-1}(K) \cdot \vol_{d-1}(\stdpolar{K}) \geq \mu_{d-1}$, where
\[
    \mu_{d-1}
        ~ = ~ \left(\frac{\pi}{2e}\right)^{\kern-2pt d-1}  \frac{d^{d}}{(d-1)!^2}.
\]
\end{lem}
%-----------------------------------------------------------------------

\begin{figure}[htbp]
    \centerline{\includegraphics[scale=0.4]{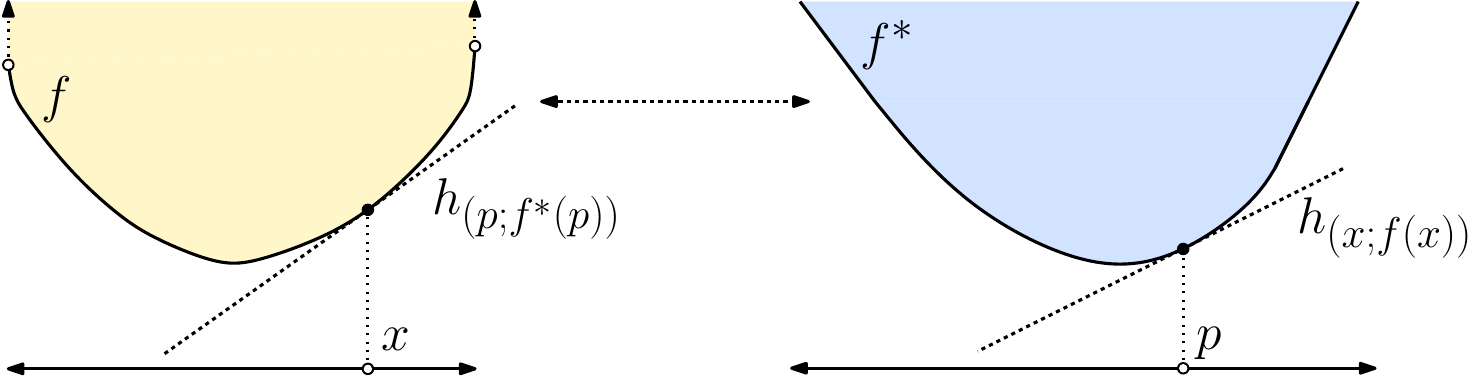}}
    \caption{The dual conjugate.}
    \label{fig:dual-conj}
\end{figure}

Next, let us consider the dual conjugate of a convex function, also known as the Legendre--Fenchel transformation. Given a lower semicontinuous convex function $f$ on $\RE^{d-1}$, its \emph{dual conjugate} $f^*$ is defined by 
\[
    f^*(p) 
        ~ = ~ \sup_{x \in \dom f} \left\{ \inner{p}{x} - f(x) \right\}
\]
(see Figure~\ref{fig:dual-conj}). 

%-----------------------------------------------------------------------
%-----------------------------------------------------------------------

This function $f^*$ is a lower semicontinuous convex function on $\RE^{d-1}$. The transformation is an involution, that is, $f^{**} = f$. A straightforward consequence of the definition is the \emph{Fenchel-Young inequality}, which states that for all $x, p \in \RE^{d-1}$, $\inner{p}{x} \leq f(x) + f^*(p)$. The next lemma encapsulates the essence of the relationship between $f$ and $f^*$. Assertions (i)--(iii) were proved in~\cite[Theorem~23.5]{Roc97}, and~(iv) and~(v) follow directly from the Fenchel--Young inequality. (Recall the notation definitions from Section~\ref{sec:notation}.)

%-----------------------------------------------------------------------
\begin{lem} \label{lem:legendre-props}
Let $f$ be a lower semicontinuous convex function on $\RE^{d-1}$, and let $x$ belong to the interior of $\dom f$. The following conditions on $p \in \RE^{d-1}$ are equivalent:
\begin{enumerate}\setlength{\itemsep}{-0.5ex}\setlength{\parsep}{0pt}
\item[$(i)$] $p \in \subgrad f(x)$

\item[$(ii)$] $x \in \subgrad f^*(p)$

\item[$(iii)$] $\inner{p}{x} = f(x) + f^*(p)$

\item[$(iv)$] $h_{(p; f^*(p))}$ supports $\epi f$ at $(x; f(x))$

\item[$(v)$] $h_{(x; f(x))}$ supports $\epi f^*$ at $(p; f^*(p))$.
\end{enumerate}
\end{lem}
%-----------------------------------------------------------------------

If $f$ satisfies \hyperref[def:reg-assump]{Regularity Assumption}~(i) ($(\interior(\dom f), f)$ is of Legendre type), then, as shown in~\cite[Theorem~26.5]{Roc97}, $(\RE^{d-1}, f^*)$ is also of Legendre type. The subdifferential is a diffeomorphism between $\dom f$ and $\RE^{d-1}$. Thus, for any $x \in \interior(\dom f)$ and $p = \Gradient f(x)$, the pair $(x,p)$ satisfies all the properties of Lemma~\ref{lem:legendre-props}. Symmetrically, this holds for any $p \in \RE^{d-1}$ and $x = \Gradient f^*(p)$. We refer to such pairs $x$ and $p$ as \emph{dual counterparts}.

\hyperref[def:reg-assump]{Regularity Assumption}~(i) allows us to simplify our notation for caps and cap bases. Since $p = \Gradient f(x)$, we can eliminate the subgradient parameter $p$ from the definitions by defining $\SCap(f, x) = \SCap(f, x, \Gradient f(x))$, and we can do the same for $\VCap$, $\DBase$, and so on.

%=======================================================================
\subsection{Dual Caps and Caps in the Dual} \label{sec:cap}
%=======================================================================

In Section~\ref{sec:func-approx}, we demonstrated the relevance of dual caps to approximation. In this section, we show that dual caps of $f$ are naturally associated with caps in the Legendre dual. Throughout, let $f$ denote a lower semicontinuous convex function on $\RE^{d-1}$ that satisfies \hyperref[def:reg-assump]{Regularity Assumption}~(i), and let $\eps$ denote a fixed approximation parameter.

\begin{figure}[htbp]
    \centerline{\includegraphics[scale=0.4,page=4]{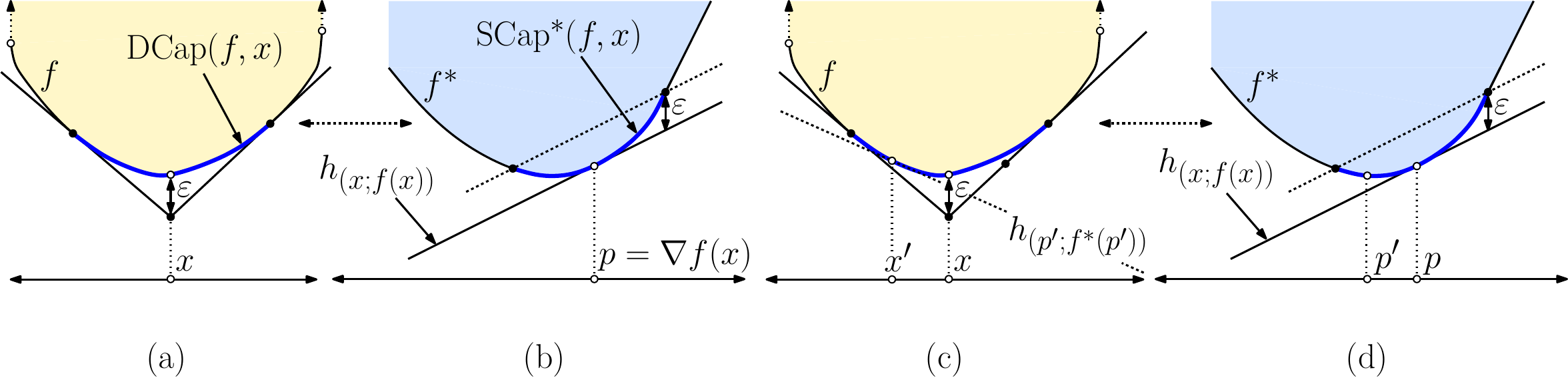}}
    \caption{A dual cap, the corresponding cap in the dual, and the proof of Lemma~\ref{lem:equiv}.}
    \label{fig:cap-dual-cap-3}
\end{figure}

Given $x \in \interior(\dom f)$, let $p = \Gradient f(x)$ be its dual counterpart. By Lemma~\ref{lem:legendre-props}(v), $h_{(x; f(x))}$ supports $\epi f^*$ at $(p; f^*(p))$. Let $H_{(x; f(x))}$ denote the closed lower halfspace bounded by $h_{(x; f(x))}$, and recall that the upward vertical translate of this halfspace by $\eps$ defines a cap of $f^*$, which is easily seen to be $\SCap(f^*, p)$ (see Figure~\ref{fig:cap-dual-cap-3}). We can think of this ``cap in the dual'' as a counterpart to the dual cap $\DCap(f, x)$. Define
\[
    \SCapD(f, x)
        ~ = ~ \SCap(f^*, \Gradient f(x)),
\]
and define $\VCapD(f, x)$ analogously using $\VCap$. This correspondence is shown in the following lemma. 

%-----------------------------------------------------------------------
%-----------------------------------------------------------------------

%-----------------------------------------------------------------------
\begin{lem} \label{lem:equiv}
Consider $\eps > 0$ and a function $f$ satisfying \hyperref[def:reg-assump]{Regularity Assumption}~(i). Given $x, x' \in \interior(\dom f)$,
\[
    x' \in \DCap(f, x)^{\downarrow} \quad\iff\quad \Gradient f(x') \in \SCapD(f, x)^{\downarrow}.
\]
\end{lem}
%-----------------------------------------------------------------------

%-----------------------------------------------------------------------
\begin{proof}
Let $p = \Gradient f(x)$ and $p' = \Gradient f(x')$ be the dual counterparts of $x$ and $x'$, respectively. By definition, $\DCap(f, x)^{\downarrow}$ consists of points $x'$ such that $(x; f(x))$ lies at vertical distance at most $\eps$ above the supporting hyperplane at $(x'; f(x'))$, which by Lemma~\ref{lem:legendre-props}(iv) is $h_{(p'; f^*(p'))}$ (see Figure~\ref{fig:cap-dual-cap-3}(c)). Similarly, $\SCap(f^*, p)^{\downarrow}$ consists of points $p'$ such that $(p'; f^*(p'))$ lies at vertical distance at most $\eps$ above the supporting hyperplane at $(p; f^*(p))$, which by Lemma~\ref{lem:legendre-props}(v) is $h_{(x; f(x))}$ (see Figure~\ref{fig:cap-dual-cap-3}(d)). Thus, we have
\begin{align*}
    x' ~ \in ~ \DCap(f, x)^{\downarrow} 
        & ~ \iff ~ f(x) - (\inner{x}{p'} - f^*(p')) ~ \leq ~ \eps \\
        & ~ \iff ~ f^*(p') - (\inner{p'}{x} - f(x)) ~ \leq ~ \eps \\
        & ~ \iff ~ p' ~ \in ~ \SCap(f^*, p)^{\downarrow} ~ = ~ \SCapD(f, x)^{\downarrow},
\end{align*}
as desired.
\end{proof}
%-----------------------------------------------------------------------

This allows us to prove the main result of this section, which states that the vertical projections of a dual base and its corresponding cap are polars of each other, subject to an appropriate translation and uniform scaling.

%-----------------------------------------------------------------------
\begin{lem} \label{lem:dual-bases}
Consider $\eps > 0$ and a function $f$ satisfying \hyperref[def:reg-assump]{Regularity Assumption}~(i). Given $x \in \interior(\dom f)$,
\[
    \DBase(f, x)^{\downarrow} - x 
        ~ = ~ \eps \SP \stdpolar{\big( \SCapD(f, x)^{\downarrow} - \Gradient f(x) \big)}.
\]
\end{lem}
%-----------------------------------------------------------------------

%-----------------------------------------------------------------------
\begin{proof}
Let $p = \Gradient f(x)$ be the dual counterpart of $x$. By the definition of the polar (Section~\ref{sec:duality}), it suffices to show that $x' \in \DBase(f,x)^{\downarrow}$ if and only if $\inner{x' - x}{p'' - p} \leq \eps$, for all $p'' \in \SCapD(f,x)^{\downarrow}$. The remainder of the proof is devoted to establishing this equivalence.

To prove the ``only if'' direction, let $x' \in \DBase(f,x)^{\downarrow}$ and $p'' \in \SCapD(f,x)^{\downarrow} = \SCap(f^*,p)^{\downarrow}$ (see Figure~\ref{fig:cap-dual-cap-5}). Since $\inner{x' - x}{p'' - p}$ is maximized when $p''$ lies on the boundary of $\SCap(f^*,p)^{\downarrow}$, we may assume that $p''$ lies on this boundary.

%-----------------------------------------------------------------------
\begin{figure}[htbp]
    \centerline{\includegraphics[scale=0.4,page=3]{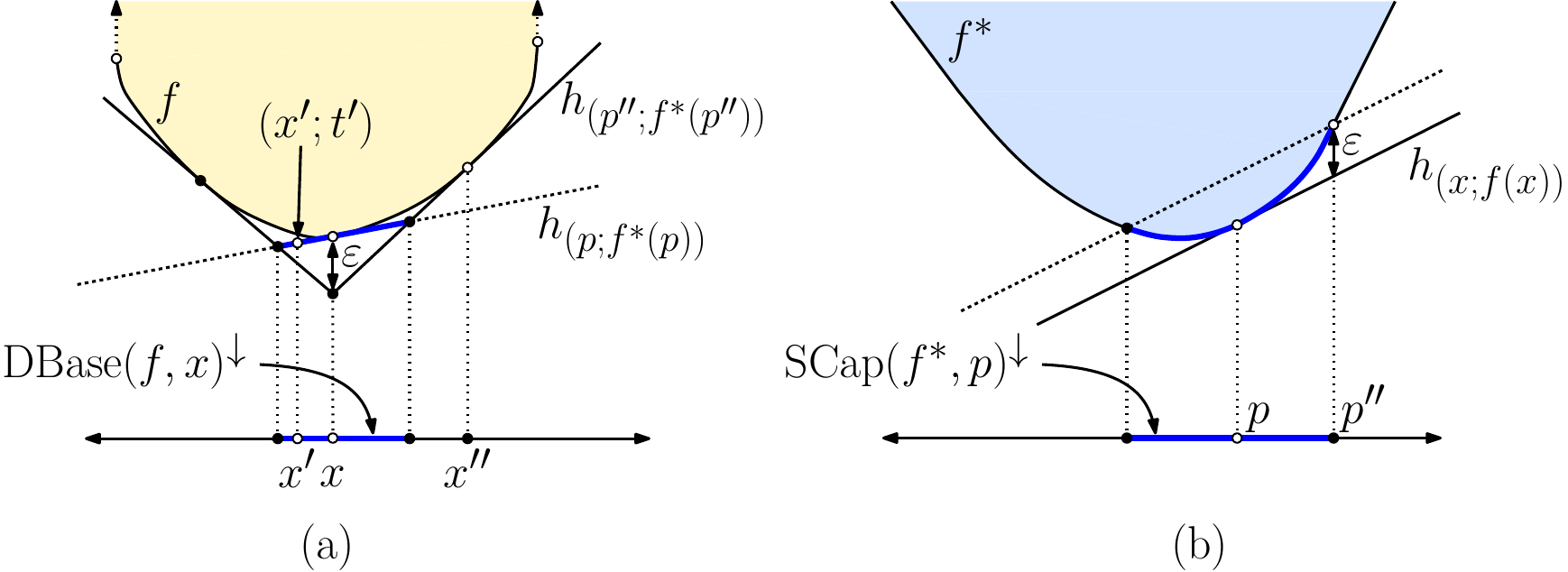}}
    \caption{Proof of Lemma~\ref{lem:dual-bases}.}
    \label{fig:cap-dual-cap-5}
\end{figure}
%-----------------------------------------------------------------------

By definition of the cap and the fact that $p''$ is on its boundary, $(p'';f^*(p''))$ is at vertical distance $\eps$ above the supporting hyperplane of $\epi f^*$ at $(p;f^*(p))$. By Lemma~\ref{lem:legendre-props}(v), this hyperplane is $h_{(x; f(x))}$, and hence
\begin{equation} \label{eqn:dual-bases-1}
    f^*(p'') - (\inner{x}{p''} - f(x))
        ~ = ~  \eps.  
\end{equation}
Similarly, since $x' \in \DBase(f,x)^{\downarrow}$, there exists $t' \in \RE$ such that $(x';t')$ lies on the supporting hyperplane to $\epi f$ at $(x;f(x))$. By Lemma~\ref{lem:legendre-props}(iv), this hyperplane is $h_{(p; f^*(p))}$, and so
\begin{equation} \label{eqn:dual-bases-2}
    t' 
        ~ = ~ \inner{p}{x'} - f^*(p).
\end{equation}

Let $x''$ denote the dual counterpart of $p''$ (see Figure~\ref{fig:cap-dual-cap-5}(a)). Clearly, $x''$ lies on the boundary of $\DCap(f, x)$, which implies that the supporting hyperplane to $\epi f$ at $(x'';f(x''))$ passes through $(x;f(x)-\eps)$. By Lemma~\ref{lem:legendre-props}(iv), this hyperplane is $h_{(p'';f^*(p''))}$. Clearly, $(x';t')$ lies within the cone, $\Cone(f, x)$, associated with the dual cap, and the hyperplane $h_{(p'';f^*(p''))}$ bounds this cone. Therefore, $(x';t')$ lies on or above this hyperplane, that is,
\[
    t' 
        ~ =    ~ \inner{p}{x'} - f^*(p) 
        ~ \geq ~ \inner{p''}{x'} - f^*(p''),
\]
or equivalently,
\begin{equation} \label{eqn:dual-bases-4}
    \inner{x'}{p''} - \inner{x'}{p} 
        ~ \leq ~ f^*(p'') - f^*(p).
\end{equation}
Thus, we have
\begin{align*}
    \inner{x'-x}{p''-p} 
        & ~ =    ~ \left( \inner{x'}{p''} - \inner{x'}{p} \right) - \inner{x}{p''} + \inner{x}{p} \\
        & ~ \leq ~ \left( f^*(p'') - f^*(p) \right) - \inner{x}{p''} + \inner{x}{p} 
                    & \text{(by Eq.~\eqref{eqn:dual-bases-4})}\\
        & ~ =    ~ \left( \inner{x}{p''} - f(x) + \eps \right) - f^*(p) -  \inner{x}{p''} + \inner{x}{p} 
                    & \text{(by Eq.~\eqref{eqn:dual-bases-1})} \\
        & ~ =    ~ \eps - (f(x) + f^*(p) - \inner{x}{p})
          ~ =    ~ \eps.
                    & \text{(by Lemma~\ref{lem:legendre-props}(iii))}
\end{align*}
This completes the ``only if'' direction.

To show the ``if'' direction, assume that for all $p'' \in \SCap(f^*,p)^{\downarrow}$, $\inner{x' - x}{p'' - p} \leq \eps$. As before, it suffices to restrict attention to $p''$ on the boundary of $\SCap(f^*,p)^{\downarrow}$, implying that Eq.~\eqref{eqn:dual-bases-1} holds. Recall $t'$ defined above and the dual counterpart $x''$ to $p''$. To show that $x' \in \DBase(f,x)^{\downarrow}$, it suffices to verify that $(x'; t')$ lies on or above the supporting hyperplane to $\epi f$ at $(x''; f(x''))$, namely $h_{(p'';f^*(p''))}$ (see Figure~\ref{fig:cap-dual-cap-5}(a)).

Since $\inner{x' - x}{p'' - p} \leq \eps$, we have
\[
    \inner{p''}{x'} - \inner{p}{x'} 
        ~ \leq ~ \eps  + \inner{x}{p''} - \inner{x}{p}.
\]
We can rewrite the right side as
\begin{align*}
    \eps + \inner{x}{p''} - \inner{x}{p} 
        & ~ = ~ \left( f^*(p'') + f(x) \right) - \inner{x}{p} 
                    & \text{(by Eq.~\eqref{eqn:dual-bases-1})} \\
        & ~ = ~ \left( f^*(p'') + \left( \inner{x}{p} - f^*(p) \right) \right) - \inner{x}{p}
                    & \text{(by Lemma~\ref{lem:legendre-props}(iii))} \\
        & ~ = ~ f^*(p'') - f^*(p).
\end{align*}
This yields $\inner{p''}{x'} - \inner{p}{x'} \leq f^*(p'') - f^*(p)$. By rearranging terms and applying Eq.~\eqref{eqn:dual-bases-2} we obtain
\[
    \inner{p''}{x'} - f^*(p'') 
        ~ \leq ~ \inner{p}{x'} - f^*(p) 
        ~ =    ~ t',
\]
which is equivalent to saying that $(x'; t')$ lies on or above $h_{(p'';f^*(p''))}$, for all $p''$ on the boundary of $\SCap(f^*,p)^{\downarrow}$, as desired.
\end{proof}
%-----------------------------------------------------------------------

%=======================================================================
\subsection{Additional Properties} \label{sec:add-prop}
%=======================================================================

Before describing our approximation constructions, in this section we present a couple of additional properties of caps, which will be used later in Section~\ref{sec:hit-approx}. The first states that the vertical projection of the base of any useful dual cap contains a ball whose radius is proportional to $\eps$. The second shows that the vertical projection of the associated cap in the dual is contained within a ball of constant radius. 

%-----------------------------------------------------------------------
\begin{lem} \label{lem:dual-boundary}
Consider a function $f$ that satisfies the \hyperref[def:reg-assump]{Regularity Assumptions} with respect to $D$, $\eps$, and $\lambda$, and let $x \in D$. Then $\DBase(f, x)^{\downarrow}$ contains a ball of radius $\eps/(1 + 2\lambda)$ centered at $x$.
\end{lem}
%-----------------------------------------------------------------------

%-----------------------------------------------------------------------
\begin{proof}
Consider any point $(s;t)$ on the boundary of $\DBase(f, x)$ (see Figure~\ref{fig:dual-boundary-2}). Since this point lies on $\DBase(f, x)$, a ray shot from $(x; f(x) - \eps)$ through this point hits $\graph f$ at some point $(x'; f(x'))$, where $x' \in \interior(\dom f)$. Consider the vertical plane passing through $x$ and $s$. It suffices to consider the restriction of $f$ to the $1$-dimensional function along this slice. Let us treat $x$ and $s$ as points on the real line. We may assume without loss of generality that $x < s < x'$. It suffices to show that $s - x \geq \eps/(1 + 2\lambda)$.

%-----------------------------------------------------------------------
\begin{figure}[htbp]
    \centerline{\includegraphics[scale=0.4,page=2]{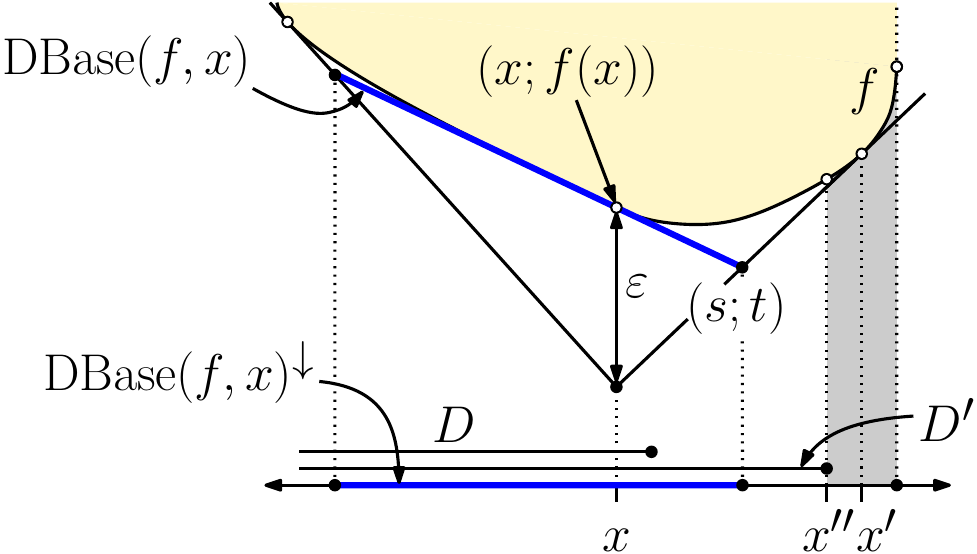}}
    \caption{Proof of Lemma~\ref{lem:dual-boundary}.}
    \label{fig:dual-boundary-2}
\end{figure}
%-----------------------------------------------------------------------

Let $f'$ denote the derivative of $f$. The supporting line to $\epi f$ at $(x; f(x))$ which has slope $f'(x)$ passes through $(s;t)$. The supporting line at $(x'; f(x'))$, which has slope $f'(x')$, passes through both $(x; f(x) - \eps)$ and $(s;t)$. As a result, we have the following.
\begin{align}
	t 
		& ~ =    ~ f'(x) (s - x) + f(x) \label{eq:dual-boundary-2} \\
    t
        & ~ =    ~ f'(x') (s - x) + f(x) - \eps \label{eq:dual-boundary-3}
\end{align}
Combining these yields
\begin{equation}
	\eps
		~ = ~ \left( f'(x') - f'(x) \right) (s - x). \label{eq:dual-boundary-5}
\end{equation}

Let $D' = D \oplus \eps B^1_2$. By \hyperref[def:reg-assump]{Regularity Assumption}~(ii), $D \subseteq D' \subseteq \interior(\dom f)$. If both $x$ and $x'$ lie within $D'$, then by \hyperref[def:reg-assump]{Regularity Assumption}~(iii), $|f'(x)|$ and $|f'(x')|$ are both bounded above by $\lambda$. By Eq.~\eqref{eq:dual-boundary-5}, we have $\eps \leq 2 \lambda (s - x)$, which implies the desired lower bound on $s - x$.

Otherwise, since $x \in D$, the interval $[x, x']$ intersects the boundary of $D'$ at a unique point, denoted $x''$. Observe that $x'' - x \geq \eps$. \hyperref[def:reg-assump]{Regularity Assumption}~(iii) implies that $f$ has Lipschitz constant $\lambda$ throughout $[x, x'']$. It follows from Lemma~\ref{lem:lipschitz} that
\begin{equation}
    f(x'') - f(x)
        ~ \leq ~ \lambda \cdot (x'' - x). \label{eq:dual-boundary-1}
\end{equation}
The point $(x''; f(x''))$ lies on or above the supporting line at $(x'; f(x'))$ (alluded to in Eq.~\eqref{eq:dual-boundary-3}), which implies that
\[
	f(x'')
		~ \geq ~ f'(x') (x'' - x) + f(x) - \eps.
\]
By combining this with Eq.~\eqref{eq:dual-boundary-1} and recalling that $x'' - x \geq \eps$, we have
\[
	f'(x')
		~ \leq ~ \frac{f(x'') - f(x) + \eps}{x'' - x}
		~ \leq ~ \frac{\lambda \cdot (x'' - x) + (x'' - x)}{x'' - x}
		~ =    ~ 1 + \lambda.
\]
Since $x \in D$, $|f'(x)| \leq \lambda$. By Eq.~\eqref{eq:dual-boundary-5} together with these bounds on $f'(x')$ and $f'(x)$, we have 
\begin{equation}
	\eps
		~ = ~ (f'(x') - f'(x)) (s - x)
		~ \leq ~ ((1 + \lambda) + \lambda) (s - x)
		~ =    ~ (1 + 2 \lambda) (s - x).
\end{equation}
Therefore $s - x \geq \eps/(1 + 2 \lambda)$, as desired.
\end{proof}
%-----------------------------------------------------------------------

We can use the polar relationship between the projections of caps and dual caps from Lemma~\ref{lem:dual-bases} to provide an upper bound on the size of a projected cap in the dual.

%-----------------------------------------------------------------------
\begin{lem} \label{lem:cap-containment}
Consider a function $f$ that satisfies the \hyperref[def:reg-assump]{Regularity Assumptions} with respect to $D$, $\eps$, and $\lambda$, and let $x \in D$. Then $\SCapD(f,x)^{\downarrow}$ is contained within a ball of radius $1 + 2\lambda$ centered at $\Gradient f(x)$.
\end{lem}
%-----------------------------------------------------------------------

%-----------------------------------------------------------------------
\begin{proof}
Let $p = \Gradient f(x)$ denote the dual counterpart of $x$. By Lemma~\ref{lem:dual-boundary}, $\DBase(f,x)^{\downarrow}$ contains a ball of radius $\eps/(1 + 2\lambda)$ centered at $x$, which implies that $(\DBase(f,x)^{\downarrow} - x)/\eps$ contains a ball of radius $1/(1+2\lambda)$ centered at the origin. By definition, $\SCapD(f,x) = \SCap(f^*,p)$. By Lemma~\ref{lem:dual-bases}, we have
\[
    (\DBase(f,x)^{\downarrow} - x)/\eps 
        ~ = ~ \stdpolar{(\SCap(f^*,p)^{\downarrow} - p)}
        ~ = ~ \stdpolar{(\SCapD(f,x)^{\downarrow} - p)}. 
\]
Due to the reciprocal nature of the polar transformation, it follows that $\SCapD(f,x)^{\downarrow} - p$ is contained within a ball of radius $1 + 2\lambda$ centered at the origin. Therefore, $\SCapD(f,x)^{\downarrow}$ is contained within a ball of radius $1 + 2\lambda$ centered at $p$, as desired.
\end{proof}
%-----------------------------------------------------------------------

%=======================================================================
\section{Hitting Sets and Approximation} \label{sec:hit-approx}
%=======================================================================

Armed with the tools developed in the previous section, in this section we will present the constructions to establish Theorem~\ref{thm:reg-func-approx}. Let us start with a high-level description of how this is done. Recall that we are given a convex function $f$ that satisfies the \hyperref[def:reg-assump]{Regularity Assumptions} with respect to a convex set $D$, and positive scalars $\eps$ and $\lambda$. By Lemma~\ref{lem:dual-outer}, it suffices to show the existence of a hitting set for all the useful $\eps$-dual caps of $f$. Specifically, we seek a discrete set $\mathcal{H} \subset \interior(\dom f)$ such that $\DCap(f, x)^{\downarrow} \cap \mathcal{H} \neq \emptyset$ for every useful $x$ (i.e., for every $x \in D$). Our approach to constructing hitting sets for any collection of dual caps involves constructing hitting sets for a related set of caps. There are two ways in which these caps arise, depending on whether they involve $f$ or $f^*$. The following lemma establishes these two distinct approaches. The first part holds by the fact that $\DBase(f,x)^{\downarrow} \subseteq \DCap(f,x)^{\downarrow}$, for all $x \in D$. The second part is a direct consequence of Lemma~\ref{lem:equiv}.

%-----------------------------------------------------------------------
\begin{lem} \label{lem:hitting-caps}
Consider a function $f$ that satisfies the \hyperref[def:reg-assump]{Regularity Assumptions} with respect to a convex body $D$, and let $X$ be a subset of $D$.
\begin{enumerate}
\item[$(i)$] Let $\mathcal{H}$ be a hitting set for the family of caps $\mathcal{C} = \{\SCap(f, x) \ST x \in X\}$, satisfying the property that for any $x \in X$, there is a point $x' \in \mathcal{H}$ such that $x' \in \DBase(f, x)^{\downarrow}$. Then $\mathcal{H}$ is a hitting set for the family of dual caps $\mathcal{C'} = \{ \DCap(f, x) \ST x \in X \}$.

\item[$(ii)$] Let $\mathcal{H}$ be a hitting set for the family of caps in the dual, $\mathcal{C} = \{\SCapD(f, x) \ST x \in X\}$. Let $\mathcal{H'}$ denote the set consisting of the dual counterparts of the points of $\mathcal{H}$. Then $\mathcal{H'}$ is a hitting set for the family of dual caps $\mathcal{C'} = \{\DCap(f, x) \ST x \in X\}$.
\end{enumerate}
\end{lem}
%-----------------------------------------------------------------------

To construct these hitting sets for caps, we will employ a classical structure from the study of convex bodies, called Macbeath regions (presented in Section~\ref{sec:macbeath} below). To generate the hitting set, we construct Macbeath regions along the lower boundary of the epigraph of the function of interest (either $f$ or $f^*$), select a constant number of points from each region and take the vertical projections of these points. 

This Macbeath-based approach will be most efficient when the caps being hit are sufficiently large. To make this notion precise, let 
\[
    t_0 
        ~ = ~ \sqrt{\vol_{d-1}(D)} \cdot \left( \frac{\eps}{\max(1,\lambda)} \right)^{\frac{d-1}{2}},
\]
and define $X_1 \subseteq D$ to be
\[
    X_1
        ~ = ~ \left\{ x \in D \ST \vol_{d-1}(\DBase(f, x)^{\downarrow}) \geq t_0 \right\}.
\]
The resulting set of \emph{large} dual caps will be handled by a hitting set derived from Lemma~\ref{lem:hitting-caps}(i). This will be presented in Section~\ref{sec:large-primal-caps}. The remaining \emph{small} dual caps are associated with the complement set $X_2 = D \setminus X_1$. These will be handled by a hitting set derived from Lemma~\ref{lem:hitting-caps}(ii). We will exploit the polar relationship between dual caps and caps in the dual (Lemma~\ref{lem:dual-bases}) together with the Mahler-volume bound (Lemma~\ref{lem:mahler-bounds}) to show that if $x \in X_2$, the associated cap in the dual, $\SCapD(f, x)$, is sufficiently large. This is made more precise in the following lemma.

%-----------------------------------------------------------------------
\begin{lem} \label{lem:dual-sizes}
Given a function $f$ that satisfies the \hyperref[def:reg-assump]{Regularity Assumptions} and a point $x \in X_2$, the associated cap in the dual, $\SCapD(f,x)$, satisfies
\[
    \vol_{d-1}(\SCapD(f, x)^{\downarrow}) 
        ~ \geq ~ \mu_{d-1} \frac{\eps^{d-1}}{t_0},
\]
where $\mu_{d-1}$ is the dimension-dependent constant of Lemma~\ref{lem:mahler-bounds}.
\end{lem}
%-----------------------------------------------------------------------

%-----------------------------------------------------------------------
\begin{proof}
Let $\DBase(f, x)$ denote the base of $x$'s dual cap. Let $p$ be $x$'s dual counterpart, so that $\SCapD(f, x) = \SCap(f^*, p)$. To simplify notation, let $C = \SCap(f^*, p)^{\downarrow}$. By Lemma~\ref{lem:dual-bases}, we have $\DBase(f, x)^{\downarrow} - x = \eps \SP \stdpolar{(C - p)}$. Noting that these are sets in $\RE^{d-1}$, scaling by a factor of $\eps$ alters the area by a factor of $\eps^{d-1}$. Thus,
\[
    \vol_{d-1}\big(\DBase(f, x)^{\downarrow}\big)
	~ = ~ \vol_{d-1}\left(\eps \SP \stdpolar{(C - p)}\right)
	~ = ~ \eps^{d-1} \cdot \vol_{d-1}\left(\stdpolar{(C - p)}\right).
\]
Since $x \in X_2$, $\vol_{d-1}(\DBase(f, x)^{\downarrow}) < t_0$, and hence $\vol_{d-1}(\stdpolar{(C-p)}) < t_0/\eps^{d-1}$. By Lemma~\ref{lem:mahler-bounds},
\[
    \vol_{d-1}\left(C\right) \cdot \vol_{d-1}\left(\stdpolar{(C-p)}\right) 
        ~ \geq ~ \mu_{d-1},
\]
and therefore, $\vol_{d-1}(\SCapD(f, x)^{\downarrow}) = \vol_{d-1}(C) \geq \mu_{d-1}\eps^{d-1}/t_0$, as desired. 
\end{proof}
%-----------------------------------------------------------------------

The upshot is that we can compute a hitting set for small dual caps efficiently by applying the Macbeath-based approach to compute a hitting set for these ``large'' caps in the dual, and then apply Lemma~\ref{lem:hitting-caps}(ii) to pull these back to hit the original dual caps.

%=======================================================================
\subsection{Macbeath Regions} \label{sec:macbeath}
%=======================================================================

\begin{figure}[htbp]
    \centerline{\includegraphics[scale=0.4,page=1]{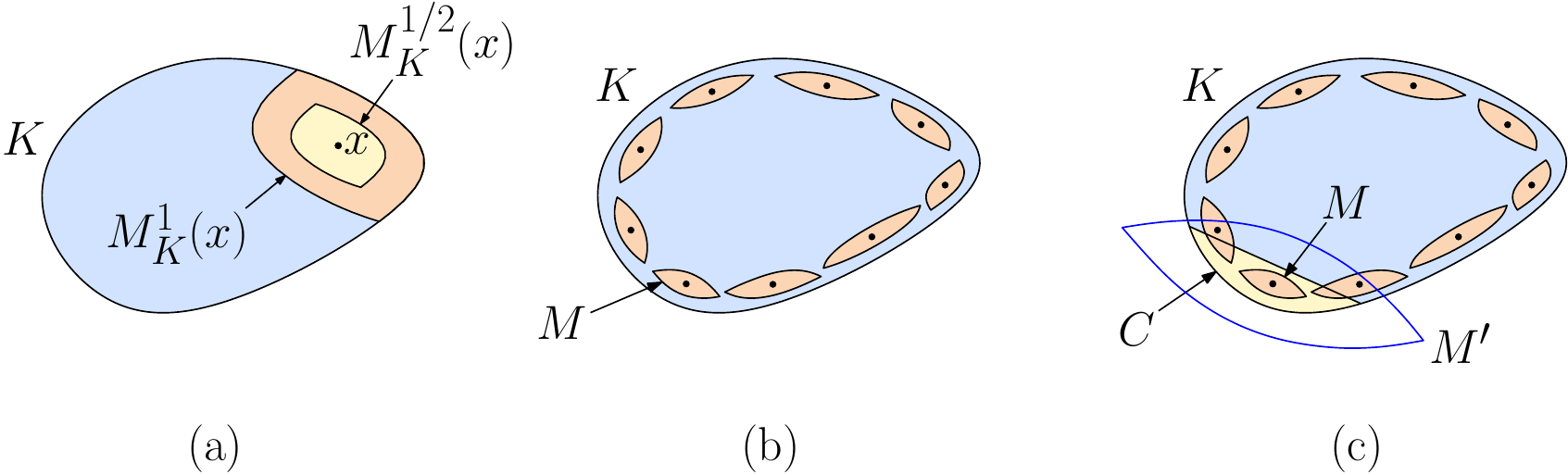}}
    \caption{Macbeath regions and Lemma~\ref{lem:cap-cover}.}
    \label{fig:macbeath}
\end{figure}

As discussed previously, our construction of hitting sets will employ a classical concept from the theory of convex sets called \emph{Macbeath regions}. Given a convex body $K$, a point $x \in K$, and a real parameter $\lambda \geq 0$, the Macbeath region $M_K^{\lambda}(x)$ (also called an \emph{$M$-region}) is defined as 
\[
    M_K^{\lambda}(x) 
        ~ = ~ x + \lambda ((K - x) \cap (x - K))
\]
(see Figure~\ref{fig:macbeath}(a)). That is, $M_K^1(x)$ is the intersection of $K$ with the body obtained by reflecting $K$ about $x$, and $M_K^{\lambda}(x)$ is a scaling of this body about $x$. Evidently, $M_K^1(x)$ is the largest centrally symmetric body centered at $x$. The scaled form is frequently used in covering and packing applications. When the body $K$ is clear from context, we will omit explicit reference to it.

%-----------------------------------------------------------------------
%-----------------------------------------------------------------------

This concept was introduced by Macbeath~\cite{Mac52}. Macbeath regions have found numerous uses in the theory of convex sets and the geometry of numbers~\cite{Bar07}, and they have been applied to several problems in the field of computational geometry, including lower bounds~\cite{BCP93, AMM09b, AMX12}, combinatorial complexity~\cite{AFM24,MuR14, AFM17c, DGJ19, AAFM22}, approximate nearest neighbor searching~\cite{AFM17a}, and computing the diameter and $\eps$-kernels~\cite{AFM17b}. 

One of these applications involves the packing and covering of a collection of caps. (In contrast to our earlier usage, we are using ``cap'' in its traditional form as the intersection of a halfspace and a convex body.) This has been extensively explored in the works of Ewald, Larman, and Rogers~\cite{ELR70}, B\'{a}r\'{a}ny and Larman~\cite{BaL88}, B\'{a}r\'{a}ny~\cite{Bar00,Bar89}, Br{\"o}nnimann {\etal}~\cite{BCP93} and Arya {\etal}~\cite{AFM17c}. We will need a variant of the covering lemma, which is presented next. The proof is a straightforward adaptation of Lemma~{3.1} in~\cite{AFM17c}, and for the sake of completeness, it is presented in Section~\ref{sec:cap-cover}.

%-----------------------------------------------------------------------
\begin{lem}[Cap Covering] \label{lem:cap-cover}
Given a convex body $K \subset \RE^d$ and any collection $\mathcal{C}$ of caps of $K$, there exist two collections of convex bodies, $\mathcal{M}$ and $\mathcal{M'}$, such that the bodies of $\mathcal{M}$ are contained within $K$ and are pairwise disjoint (see Figure~\ref{fig:macbeath}(b)). Each $M \in \mathcal{M}$ is associated with a corresponding body in $\mathcal{M'}$, denoted $M'$, such that $M \subseteq M'$. $M'$ is called $M$'s \emph{expanded body}. These sets satisfy the following:
\begin{enumerate} \setlength{\itemsep}{-0.5ex}\setlength{\parsep}{0pt}%
\item[$(i)$] For all $M \in \mathcal{M}$, $\vol_d(M') = c \cdot \vol_d(M)$, for some constant $c$ depending on the dimension.

\item[$(ii)$] For any cap $C \in \mathcal{C}$, there exists $M \in \mathcal{M}$ such that $M \subseteq C \subseteq M'$, where $M'$ is $M$'s expanded body (see Figure~\ref{fig:macbeath}(c)).
\end{enumerate}
\end{lem}
%-----------------------------------------------------------------------

%=======================================================================
\subsection{Hitting Large Caps in the Primal} \label{sec:large-primal-caps}
%=======================================================================

In this section, we will explain how to apply the Macbeath-region machinery to construct a hitting set for large caps in $f$. The approach is to invoke Lemma~\ref{lem:cap-cover} to construct a collection of regions that cover the lower portion of $\epi f$, sample a constant number of points from each region and then take the vertical projections of these points to form the hitting set.

We bound the number of points by bounding the number of Macbeath regions. We will do this by recalling Lemma~\ref{lem:cap-ratio}, which established the existence of an ellipsoid close to $\graph f$ whose projection lies within the projection of the dual cap's base. This will be combined with a classical sampling technique, called $\epsilon$-nets. The key result we need is encapsulated in the following lemma.

%-----------------------------------------------------------------------
\begin{lem}[Epsilon-Nets for Ellipsoids] \label{lem:ellipse-net}
Given a convex body $\Omega$ in $\RE^d$ and a constant $\gamma$, where $0 < \gamma < 1$, there exists a discrete set $N \subseteq \Omega$ whose size depends only on $d$ and $\gamma$, such that for any ellipsoid $E \subseteq \Omega$ where $\vol_d(E) \geq \gamma \cdot \vol_d(\Omega)$, $E$ contains at least one point of $N$.
\end{lem}
%-----------------------------------------------------------------------

%-----------------------------------------------------------------------
\begin{proof}
Let us review some standard facts. A \emph{set system} is a pair $(X,\mathcal{F})$, where $X$ is a (possibly infinite) set and $\mathcal{F}$ is a collection of subsets of $X$. Let $\mu$ be a measure on $X$. Given a set system $(X,\mathcal{F})$ and a parameter $\epsilon > 0$ (not to be confused with the $\eps$ used for approximation), a set $N \subseteq X$ is an \emph{$\epsilon$-net} of $(X,\mathcal{F})$ if for each $F \in \mathcal{F}$ with $\mu(F) \geq \epsilon \cdot \mu(X)$, $F$ contains at least one point of $N$. 

The complexity of a set system can be described by a quantity called its \emph{VC-dimension}~\cite{AHW87, Mus22}. We need only two standard facts regarding this concept. First, any set system of constant VC-dimension has an $\epsilon$-net of size $O\big( \inv{\epsilon} \log \inv{\epsilon} \big)$~\cite{AHW87,Mat02}, and second, the set system $(\Omega,\mathcal{E})$, where $\Omega$ is a bounded convex body in $\RE^d$ endowed with the Lebesgue measure and $\mathcal{E}$ is the set of ellipsoids contained in $\Omega$ has VC-dimension at most $\binom{d+2}{d} = O(d^2)$~\cite[Proposition~{10.3.2}]{Mat02}. Setting $\epsilon = \gamma$ and observing that $\gamma$ is a constant yields the desired set $N$.
\end{proof}
%-----------------------------------------------------------------------

Our next lemma is the main result of this section. It states that there exists a small hitting set for the large dual caps.

%-----------------------------------------------------------------------
\begin{lem} \label{lem:dcap-hitting}
Consider a function $f$ that satisfies the \hyperref[def:reg-assump]{Regularity Assumptions} with respect to a compact domain $D$ and parameter $\eps > 0$. In addition, assume that the minimum width of $D$ is at least $\eps$. For any $t > 0$, define the set 
\[
    X
        ~ = ~ \left\{ x \in D \ST \vol_{d-1}(\DBase(f, x)^{\downarrow}) \geq t \right\}.
\]
Then there exists a hitting set $\mathcal{H} \subset \interior(\dom f)$ of size $O(\vol_{d-1}(D) / t)$ for the family of caps $\mathcal{C} = \{\SCap(f, x) \ST x \in X\}$, satisfying the property that for every $x \in X$, there exists a point $x' \in \mathcal{H}$ such that $x' \in \DBase(f,x)^{\downarrow}$.
\end{lem}
%-----------------------------------------------------------------------

%-----------------------------------------------------------------------
\begin{proof}
Define $\mathcal{C}' = \{\VCap(f, x) \ST x \in X\}$. By the \hyperref[def:reg-assump]{Regularity Assumptions}, $\dom f$ is bounded and $D \subseteq \interior(\dom f)$. Because $D$ is compact, it follows that $\sup_{x \in D} f(x)$ and $\sup_{x \in D} \|\Gradient(f(x))\|$ are both finite. This implies that there exists a lower horizontal halfspace such that all caps in $\mathcal{C}'$ lie inside the convex body $K$ formed by intersecting $\epi f$ with this halfspace (see Figure~\ref{fig:dcap-hitting}(a)). 

We apply Lemma~\ref{lem:cap-cover} to $K$ and $\mathcal{C}'$, obtaining collections $\mathcal{M}$ and $\mathcal{M'}$ of convex bodies. For every cap $C \in \mathcal{C}'$, there exists an $M \in \mathcal{M}$ and its associated body $M' \in \mathcal{M'}$ such that $M \subseteq C \subseteq M'$. Discard from $\mathcal{M}$ any $M$ for which no such cap exists. Since each cap in $\mathcal{C}'$ has vertical width $\eps$, the surviving bodies in $\mathcal{M}$ lie entirely within vertical distance $\eps$ of $\graph f$.

%-----------------------------------------------------------------------
\begin{figure}[htbp]
    \centerline{\includegraphics[scale=0.4]{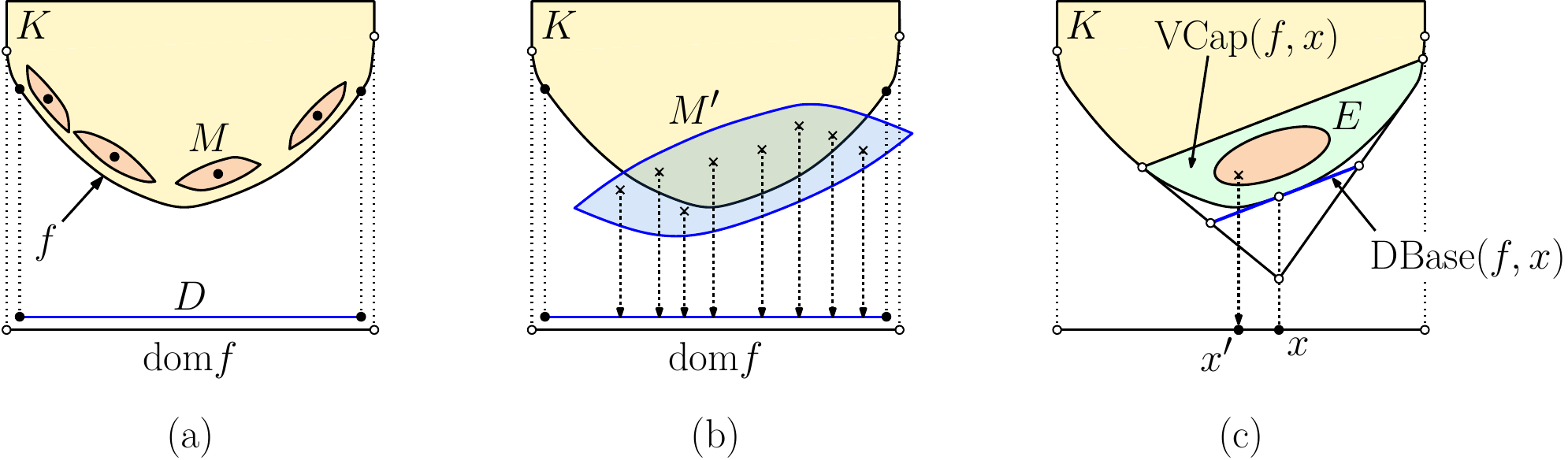}}
    \caption{Proof of Lemma~\ref{lem:dcap-hitting}.}
    \label{fig:dcap-hitting}
\end{figure}
%-----------------------------------------------------------------------

Let $c''$ be a constant whose value will be specified later. For each $M' \in \mathcal{M'}$ we invoke Lemma~\ref{lem:ellipse-net} (with $M'$ and $c''$ taking the roles of $\Omega$ and $\gamma$, respectively) to obtain a set $N \subset M'$ such that any ellipsoid contained in $M'$ with volume at least $c'' \cdot \vol(M')$ contains at least one point of $N$ (see Figure~\ref{fig:dcap-hitting}(b)).

Project each point of $N$ vertically downward onto $\RE^{d-1}$ (see Figure~\ref{fig:dcap-hitting}(b)), and add the resulting point to $\mathcal{H}$. Repeating this for all $M \in \mathcal{M}$ yields the set $\mathcal{H}$. We eliminate any points in $\mathcal{H}$ that lie outside $\dom f$.

To establish correctness, consider any point $x \in X$. By Lemma~\ref{lem:cap-ratio}, there exists an ellipsoid $E \subseteq \VCap(f,x)$ such that $E^{\downarrow} \subseteq \DBase(f,x)^{\downarrow}$, and
\[
    \vol(E) 
        ~ \geq ~ c \cdot \vol_d(\VCap(f,x))
        ~ \geq ~ c' \eps \cdot \vol_{d-1}(\DBase(f,x)^{\downarrow}),
\]
for two constants $c$ and $c'$, which depend on the dimension (see Figure~\ref{fig:dcap-hitting}(c)). By Lemma~\ref{lem:cap-cover}(ii), there exists $M \in \mathcal{M}$ such that $M \subseteq \VCap(f,x) \subseteq M'$. (Note that $M$ could not have been discarded in the construction process.) By the above inclusions and Lemma~\ref{lem:cap-cover}(i), we have 
\[
    \vol(E)
        ~ \geq ~ c \cdot \vol_d(M)
        ~ =    ~ \frac{c}{c_{\SP 0}} \vol_d(M'),
\]
where $c_{\SP 0}$ is the constant from Lemma~\ref{lem:cap-cover}(i). Setting $c'' = c/c_0$ ensures that $N$ contains a point of $E$ whose vertical projection, $x'$, will be included in $\mathcal{H}$. Since $E^{\downarrow} \subseteq \SCap(f,x)^{\downarrow}$ and $E^{\downarrow} \subseteq \DBase(f,x)^{\downarrow}$, it follows that $x'$ hits $\SCap(f,x)$ and lies in $\DBase(f,x)^{\downarrow}$, as desired.

Finally, we bound the size of $\mathcal{H}$. Consider the subset of $K$ that lies within vertical distance $\eps$ of $\graph f$. Its volume is $\eps \cdot \vol_{d-1}(\dom f)$. By the above inclusions and Lemma~\ref{lem:cap-ratio}, for each $M \in \mathcal{M}$,
\[
    \vol_d(M)
        ~ =    ~ \frac{1}{c_{\SP 0}} \vol_d(M')
        ~ \geq ~ \frac{1}{c_{\SP 0}} \vol_d(\VCap(f,x))
        ~ \geq ~ \frac{c' \eps}{c_{\SP 0} c} \cdot \vol_{d-1}(\DBase(f,x)^{\downarrow})
        ~ \geq ~ \frac{c'}{c_{\SP 0} c} \eps t.
\]
The bodies in $\mathcal{M}$ are pairwise disjoint and (after discarding) they all lie within vertical distance $\eps$ of $\graph f$. Hence, by a simple packing argument we have
\[
    |\mathcal{M}| 
        ~ = ~ O\left( \frac{\eps \cdot \vol_{d-1}(\dom f)}{\eps \SP t} \right)
	~ = ~ O\left( \frac{\vol_{d-1}(\dom f)}{t} \right).
\]

By \hyperref[def:reg-assump]{Regularity Assumption}~(ii), $\dom f \subseteq D \oplus 2\eps B^{d-1}_2$. Since $D$ has minimum width at least $\eps$, Lemma~\ref{lem:expansion} implies $\vol_{d-1}(\dom f) = O(\vol_{d-1}(D))$. Thus $|\mathcal{M}| = O(\vol_{d-1}(D)/t)$. Since each $M \in \mathcal{M}$ contributes only a constant number of points to $\mathcal{H}$, the same asymptotic bound holds for $|\mathcal{H}|$. 
\end{proof}

%=======================================================================
\subsection{Hitting Large Caps in the Dual} \label{sec:large-caps}
%=======================================================================

In this section, we consider the task of bounding the size of a hitting set for the set of large caps in the dual $f^*$. The construction is similar to the one given in the previous section, but since $\dom f^*$ covers all of $\RE^{d-1}$, some additional effort is required to restrict the size of the region to be covered. The following lemma shows how to apply the Macbeath-region machinery to hit all the useful $\eps$-caps of $U^*$ whose bases have sufficiently large area.

%-----------------------------------------------------------------------
\begin{lem} \label{lem:cap-hitting}
Consider a function $f$ that satisfies the \hyperref[def:reg-assump]{Regularity Assumptions} with respect to the domain $D$ and parameters $\eps, \lambda > 0$. For any $t > 0$, define the set 
\[
    X
        ~ = ~ \left\{ x \in D \ST \vol_{d-1}\big( \SCapD(f, x)^{\downarrow} \big) \geq t \right\}.
\]
Then there exists a hitting set $\mathcal{H} \subset \RE^{d-1}$ of size $O(\max(1,\lambda)^{d-1}/t)$ for the family of caps in the dual, $\mathcal{C} = \{\SCapD(f, x) \ST x \in X\}$.
\end{lem}
%-----------------------------------------------------------------------

%-----------------------------------------------------------------------
\begin{proof}
Consider any point $x \in X$, and let $p = \Gradient f(x)$ be its dual counterpart. By Lemma~\ref{lem:cap-containment}, the vertical projection $\SCapD(f,x)^{\downarrow}$ is contained within a ball of radius $1 + 2\lambda$ centered at $\Gradient f(x)$. By Lemma~\ref{lem:legendre-props}(i), this ball is centered at $p$. Since $x \in D$, \hyperref[def:reg-assump]{Regularity Assumption}~(iii) implies that $\|\Gradient f(x)\| \leq \lambda$, and hence $\|p\| \leq \lambda$. By the triangle inequality, $\SCap(f^*,p)^{\downarrow}$ lies within a ball of radius $(1 + 2\lambda) + \lambda = 1 + 3\lambda$ centered at the origin, which we denote by $B$. Since $X \subseteq D$, the vertical projection of every cap in $\mathcal{C}$ lies within $B$.

Define $\mathcal{C}' = \{\VCapD(f, x) \ST x \in X\}$. By the above remarks, the caps of $\mathcal{C}'$ are contained in the portion of $\epi f^*$ lying above $B$. We can convert the unbounded set $\epi f^*$ into a convex body $K$ by intersecting it with the vertical cylinder whose cross section is $B$, and then we truncate it from above by a sufficiently high horizontal hyperplane that does not intersect any cap in $\mathcal{C}'$. The existence of such a hyperplane follows from the facts that we need only cover points $p \in B$ together with the fact that $\|\Gradient f^*\|$ is bounded throughout its domain because $\dom f$ is bounded.

We apply Lemma~\ref{lem:cap-cover} to $K$ and $\mathcal{C}'$, obtaining collections $\mathcal{M}$ and $\mathcal{M'}$ of convex bodies. For every cap $C \in \mathcal{C}'$, there exists $M \in \mathcal{M}$ such that $M \subseteq C \subseteq M'$. Discard any $M \in \mathcal{M}$ for which no such cap $C$ exists. Since each cap in $\mathcal{C}'$ has vertical width $\eps$, the surviving bodies in $\mathcal{M}$ lie entirely within vertical distance $\eps$ of $\graph f^* \cap B$ (see Figure~\ref{fig:cap-hitting}(a)). For each surviving body $M$, select an arbitrary point from it (say its center), project this point vertically downward onto $\RE^{d-1}$, and add the resulting point to $\mathcal{H}$. Repeating this for all $M \in \mathcal{M}$ yields the set $\mathcal{H}$.

%-----------------------------------------------------------------------
\begin{figure}[htbp]
    \centerline{\includegraphics[scale=0.4]{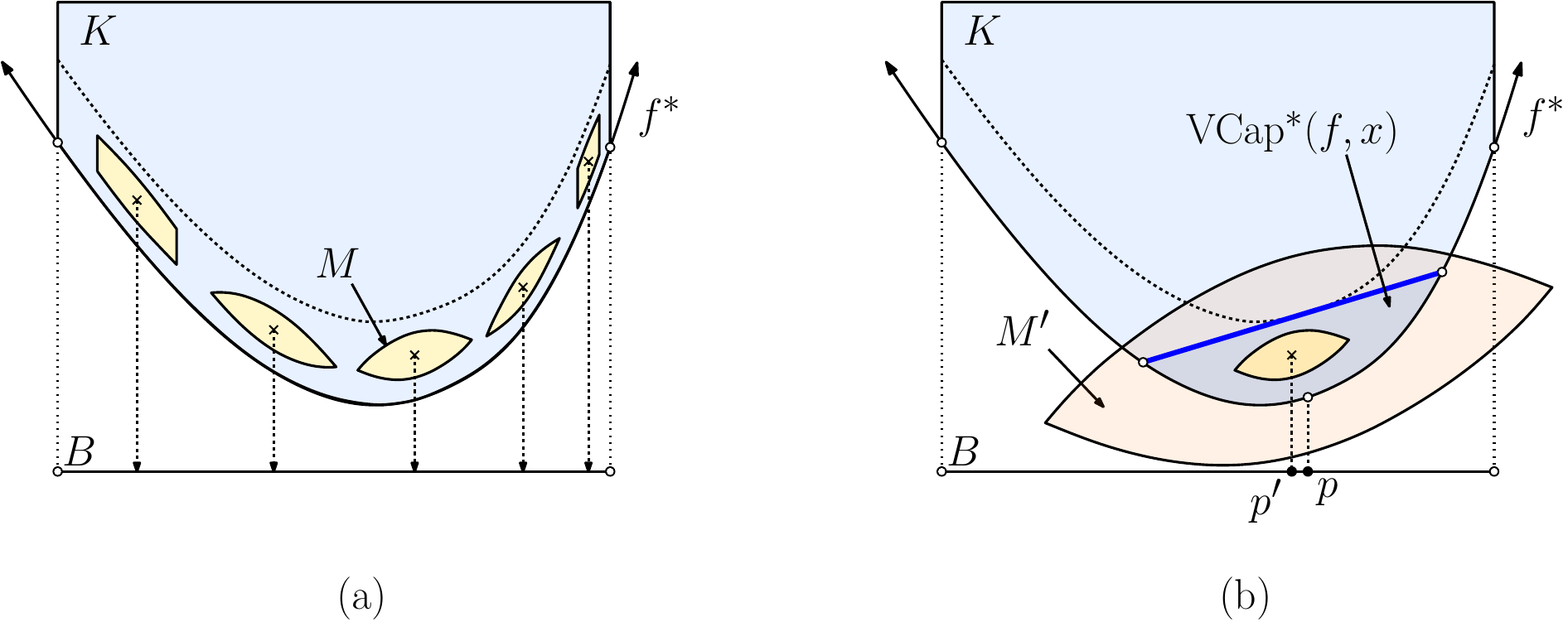}}
    \caption{Proof of Lemma~\ref{lem:cap-hitting}. (The point $p$ in (b) is the dual counterpart of $x$.)}
    \label{fig:cap-hitting}
\end{figure}
%-----------------------------------------------------------------------

To establish correctness, consider any point $x \in X$. By Lemma~\ref{lem:cap-cover}(ii), there exists an $M \in \mathcal{M}$ such that $M \subseteq \VCapD(f,x) \subseteq M'$. (Note that $M$ could not have been discarded in the construction process.) The vertical projection of some point of $M$ was included in $\mathcal{H}$. Letting $p'$ denote this point, we have 
\[
    p' 
        ~ \in ~ \VCapD(f,x)^{\downarrow} 
        ~ =   ~ \SCapD(f,x)^{\downarrow},
\]
thus, this cap is hit.

To bound the size of $\mathcal{H}$, consider the subset of $\epi f^*$ lying above $B$ and within vertical distance $\eps$ of $\graph f^*$. Its volume is $\eps \cdot \vol_{d-1}(B) = \eps \cdot (1 + 3 \lambda)^{d-1} \omega_{d-1}$, where $\omega_{d-1} = \vol_{d-1}(B^{d-1}_2)$, which is a constant depending on $d$. By the above inclusions and Lemma~\ref{lem:cap-cover}(i), for each $M \in \mathcal{M}$,
\[
    \vol_d(M)
        ~ =    ~ \frac{1}{c_{\SP 0}} \vol_d(M')
        ~ \geq ~ \frac{1}{c_{\SP 0}} \vol_d(\VCapD(f,x)),
\]
where $c_{\SP 0}$ is the constant of Lemma~\ref{lem:cap-cover}(i). Cap $\VCapD(f,x)$ contains a cone with base $\VBase(f^*,p)$ and apex $(p;f^*(p))$, whose volume is 
\[
    \frac{\eps}{d} \cdot \vol_{d-1} \left( \VBase(f^*, p)^{\downarrow} \right) 
        ~ = ~ \frac{\eps}{d} \cdot \vol_{d-1} \left( \SCapD(f,x)^{\downarrow} \right).
\]
Therefore,
\[
     \vol_d(M)
        ~ \geq ~ \frac{\eps}{c_{\SP 0} d} \cdot \vol_{d-1}(\SCapD(f,x)^{\downarrow})
        ~ \geq ~ \frac{1}{c_{\SP 0} d} \eps \SP t.
\]
The bodies in $\mathcal{M}$ are pairwise disjoint and (after discarding) lie above $B$ within vertical distance $\eps$ of $\graph f^*$. Hence, by a simple packing argument, 
\[
    |\mathcal{M}| 
        ~ = ~ O\left( \frac{\eps (1 + 3 \lambda)^{d-1}}{\eps \SP t} \right) 
        ~ = ~ O\left( \frac{(1 + 3 \lambda)^{d-1}}{t} \right).
\]
Since each $M \in \mathcal{M}$ contributes one point to $\mathcal{H}$, the same bound holds for $|\mathcal{H}|$. 
\end{proof}
%-----------------------------------------------------------------------

%=======================================================================
\subsection{Wrapping it up} \label{sec:wrap}
%=======================================================================

In this section, we combine the results of Sections~\ref{sec:large-primal-caps} and~\ref{sec:large-caps} to complete the proof of Theorem~\ref{thm:reg-func-approx}. Recall that $D$ is a compact convex domain in $\RE^{d-1}$ of minimal width at least $\eps$, and $f$ is a convex function satisfying the \hyperref[def:reg-assump]{Regularity Assumptions} with respect to $D$ and positive parameters $\eps$ and $\lambda$. Also, recall that
\begin{align*}
    t_0 
        & ~ = ~ \sqrt{\vol_{d-1}(D)} \cdot \left( \frac{\eps}{\max(1,\lambda)} \right)^{\frac{d-1}{2}} \\
    X_1
        & ~ = ~ \left\{ x \in D \ST \vol_{d-1}(\DBase(f, x)^{\downarrow}) \geq t_0 \right\} \\
    X_2
        & ~ = ~  D \setminus X_1.
\end{align*}

\begin{proof} (of Theorem~\ref{thm:reg-func-approx})

By Lemma~\ref{lem:dcap-hitting} and Lemma~\ref{lem:hitting-caps}(i), there exists a hitting set $\mathcal{H}_1$ for the family of dual caps $\{ \DCap(f, x) \ST x \in X_1 \}$ of size 
\[
    |\mathcal{H}_1|
        ~ = ~ O\left( \frac{\vol_{d-1}(D)}{t_0} \right)
        ~ = ~ O\left( \left( \frac{\max(1,\lambda)}{\eps} \right)^{\frac{d-1}{2}} \cdot \sqrt{\vol_{d-1}(D)} \right).
\]

For any $x \in X_2$, we have $\vol_{d-1}(\DBase(f, x)^{\downarrow}) < t_0$. By Lemma~\ref{lem:dual-sizes}, $\vol_{d-1}(\SCapD(f, x)^{\downarrow}) > t'_0$,
where 
\[
    t'_0 
        ~ = ~ \mu_{d-1} \frac{\eps^{d-1}}{t_0}.
\]
It follows from Lemma~\ref{lem:cap-hitting} that there exists a hitting set $\mathcal{H}_2$ for the family of caps in the dual $\{\SCapD(f, x) \ST x \in X_2\}$ of size
\[
    |\mathcal{H}_2| 
        ~ = ~ O\left( \frac{\max(1,\lambda)^{d-1}}{t'_0} \right)
        ~ = ~ O\left( \left( \frac{\max(1,\lambda)}{\eps} \right)^{d-1} \kern-5pt \cdot t_0  \right)
        ~ = ~ O\left( \left( \frac{\max(1,\lambda)}{\eps} \right)^{\frac{d-1}{2}} \kern-5pt \cdot \sqrt{\vol_{d-1}(D)}\right).
\]
By Lemma~\ref{lem:hitting-caps}(ii), $\mathcal{H}_2$ is a hitting set for the family of dual caps $\{ \DCap(f, x) \ST x \in X_2 \}$.

Therefore, the union $\mathcal{H} = \mathcal{H}_1 \cup \mathcal{H}_2$ is a hitting set for the entire family of dual caps $\{ \DCap(f, x) \ST x \in D \}$ of size 
\[
    |\mathcal{H}| 
        ~ = ~ O\left( \left( \frac{\max(1,\lambda)}{\eps} \right)^{\frac{d-1}{2}} \cdot \sqrt{\vol_{d-1}(D)}\right)
        ~ = ~ O\left( \left( \max(1,\lambda) \cdot \frac{\vrad(D)}{\eps} \right)^{\frac{d-1}{2}} \right).
\]

For each $y \in \mathcal{H}$, let $H^+(y)$ denote the closed upper halfspace defined by the supporting hyperplane to $\epi f$ at the point $(y; f(y))$. By Lemma~\ref{lem:dual-outer}, the intersection $\bigcap_{y \in \mathcal{H}} H^+(y)$ is a piecewise-linear, convex lower $\eps$-approximation to $f$ on $D$.
\end{proof}

%=======================================================================
\section{Additional Results} \label{sec:additional}
%=======================================================================

%=======================================================================
\subsection{Nonuniform Area-Based Bounds} \label{sec:nonuniform}
%=======================================================================

In this section, we present a nonuniform bound very similar to that of Theorem~\ref{thm:main}. This is derived from a result due to Gruber~\cite{Gru93a}, who showed that if $K$ is a strictly convex body and $\partial K$ is twice differentiable ($C^2$ continuous), then there exists a constant $k_d$ (depending only on the dimension $d$) and a scalar $\eps_0$ depending on $K$, such that for any $0 < \eps \leq \eps_0$, the number of bounding halfspaces needed to achieve an $\eps$-approximation to $K$ is at most
\begin{equation} \label{eq:gruber}
    k_d \left( \frac{1}{\eps} \right)^{\kern-2pt\frac{d-1}{2}} \int_{\partial K} \kappa(x)^{\frac{1}{2}} d\sigma(x), 
\end{equation}
where $\kappa$ and $\sigma$ denote the Gaussian curvature of $K$ and ordinary surface area measure, respectively. (B{\" o}r{\" o}czky showed that the requirement that $K$ be ``strictly'' convex can be eliminated~\cite{Bor00}.) Because the square root function is concave and $\int_{\partial K} d\sigma(x) = \area(K)$, we may apply Jensen's inequality to obtain
\[
    \frac{1}{\area(K)} \int_{\partial K} \kappa(x)^{\frac{1}{2}} d\sigma(x)
        ~ \leq ~ \left( \frac{1}{\area(K)} \int_{\partial K} \kappa(x) d\sigma(x) \right)^{\frac{1}{2}}.
\]
Thus,
\[
    \int_{\partial K} \kappa(x)^{\frac{1}{2}} d\sigma(x) 
        ~ \leq ~ \left(\area(K) \int_{\partial K} \kappa(x) d\sigma(x) \right)^{\frac{1}{2}}.
\]
By the Gauss--Bonnet theorem~\cite{Car76}, the total Gaussian curvature of $K$ is bounded by some quantity $\zeta_d$, depending only on $d$. Also, by definition,
\[
    \arad(K)
        ~ = ~ \left( \frac{\area(K)}{\area(B^d_2)} \right)^{\kern-2pt \frac{1}{d-1}}\kern-20pt.
\]
Therefore,
\[
    \int_{\partial K} \kappa(x)^{\frac{1}{2}} d\sigma(x) 
        ~ \leq ~ (\zeta_d \cdot \area(K))^{\frac{1}{2}}
        ~ =    ~ \left( \zeta_d \SP \area(B^d_2) \right)^{\frac{1}{2}} (\arad(K))^{\frac{d-1}{2}}.
\]
Substituting the above quantity into Eq.~\eqref{eq:gruber} and setting $c_d$ to the constant $k_d(\zeta_d \SP \area(B^d_2))^{1/2}$, we obtain the following.

%-----------------------------------------------------------------------
\begin{thm} \label{thm:nonuniform}
For any integer $d \geq 2$ and any convex body $K \subseteq \RE^d$ whose boundary is $C^2$ smooth, there exists $\eps_0$ depending on $K$, such that for any $0 < \eps \leq \eps_0$, there exists an $\eps$-approximating polytope $P$ having at most
\[
    c_d \left(\frac{\arad(K)}{\eps}\right)^{\frac{d-1}{2}}
\]
facets, where $c_d$ is a constant (depending on $d$).
\end{thm}
%-----------------------------------------------------------------------

\begin{figure}[htbp]
    \centerline{\includegraphics[scale=0.4]{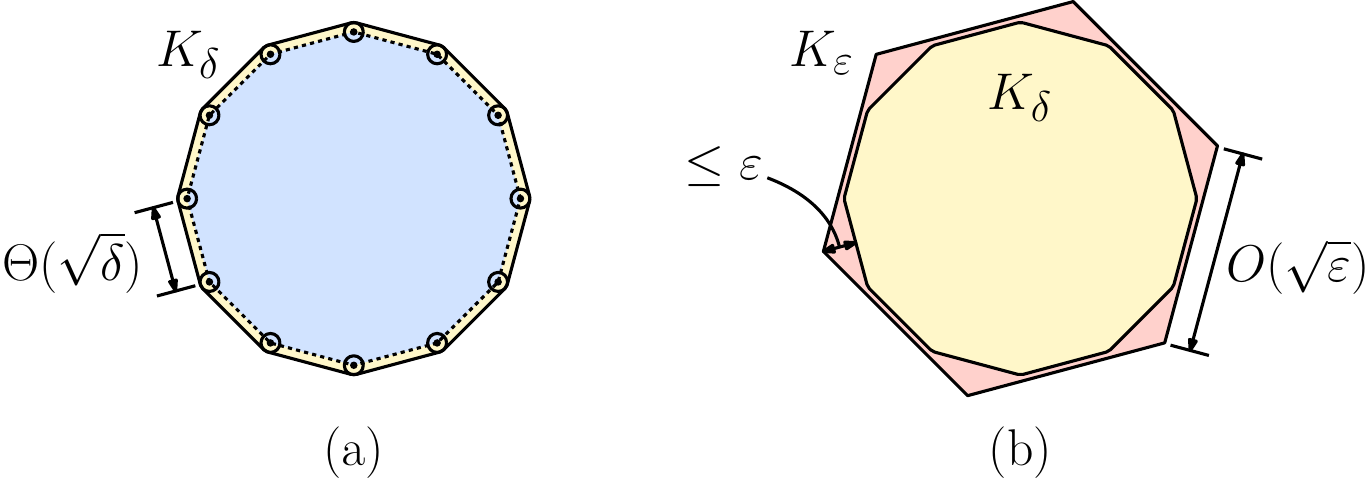}}
    \caption{Why Theorem~\ref{thm:nonuniform} cannot be used to generate a uniform bound.}
    \label{fig:counterexample}
\end{figure}

Note that the bound in this theorem matches the uniform bound of Theorem~\ref{thm:main}. However, this approach cannot be used to produce a uniform bound. To see why, suppose, to the contrary, that such a bound existed, even in $\RE^2$. That is, there exists a constant $k_2$ and positive $\eps_0$ such that for all $\eps \leq \eps_0$ and all convex bodies $K$ (of width at least $\eps$ in every direction) in $\RE^2$, there exists an $\eps$-approximating polygon whose number of sides satisfies Eq.~\eqref{eq:gruber}. Consider any $\eps \leq \min(\eps_0, 1/9)$, and let $0 < \delta \leq \eps$ be a sufficiently small value (chosen below). Set $m = \lfloor 1/\sqrt{\delta} \rfloor$, and define $K_{\delta}$ to be the Minkowski sum of a regular $m$-gon inscribed in a unit circle and the Euclidean ball of radius $\delta$ (see Figure~\ref{fig:counterexample}(a)). Observe that since $m \geq 3$ and $\delta \leq 1/9$, $K_{\delta}$ satisfies the minimum width requirements. It consists of $m$ straight edges, each of length $\Theta(\sqrt{\delta})$, connected by $m$ circular arcs, each of radius $\delta$ and subtending an angle of $2 \pi/m$. Since $\delta \le \eps$, it is straightforward to show that any convex polygon $K_{\eps}$ that $\eps$-approximates $K_{\delta}$ requires $\Omega(1/\sqrt{\eps})$ sides (see Figure~\ref{fig:counterexample}(b)). (As $\delta$ decreases relative to $\eps$, $K_{\delta}$ approaches a unit disk, and it is easy to show that in order to maintain a distance of at most $\eps$, each side can have length at most $c \sqrt{\eps}$, for some constant $c$.) 

%-----------------------------------------------------------------------
%-----------------------------------------------------------------------

Boundary points along the flat sides of $K_{\delta}$ have zero curvature, and boundary points within each circular arc have curvature $1/\delta$. Since the circular arcs together cover a distance of $2 \pi \delta$ of the boundary, it follows that 
\[
    \int_{\partial K_{\delta}} \kappa(x)^{1/2} d\sigma(x) 
        ~ = ~ \frac{2 \pi \delta}{\sqrt{\delta}}
        ~ = ~ \Theta\left(\sqrt{\delta}\right).
\]
 Therefore, the hypothesized uniform bound would imply the existence of an $\eps$-approximating polygon with $O\big(\sqrt{\delta/\eps}\big)$ sides, contradicting the lower bound of $\Omega(1/\sqrt{\eps})$ for all sufficiently small $\delta$.%
\footnote{Note that we cannot apply Gruber's or B{\" o}r{\" o}czky's theorems directly to $K_{\delta}$, since its boundary is not twice differentiable. In particular, the second derivative is discontinuous at the joints where each edge meets a circular arc. We can easily fix this by creating a sufficiently small gap at each joint and introducing a smooth polynomial spline of constant degree to fill the gap. Although the resulting body is not strictly convex, B{\" o}r{\" o}czky showed that this assumption is not necessary for the bound to hold.} 

%=======================================================================
\subsection{Proof of the Cap-Covering Lemma} \label{sec:cap-cover}
%=======================================================================

\begin{figure}[htbp]
    \centerline{\includegraphics[scale=0.4]{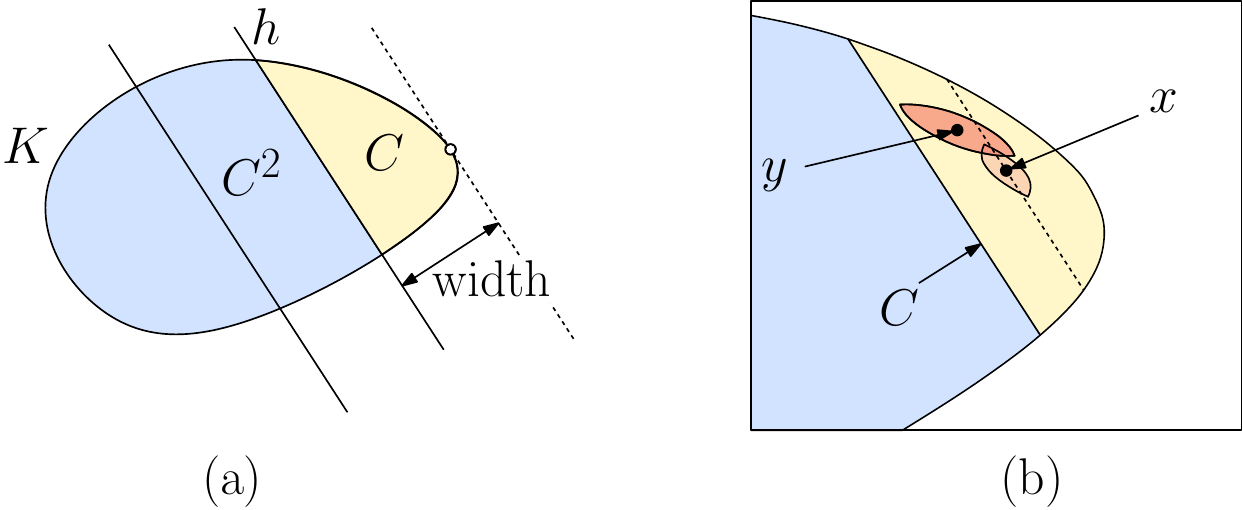}}
    \caption{Proof of the Cap-Covering lemma.}
    \label{fig:cap-cover}
\end{figure}

In this section, we present a proof of Lemma~\ref{lem:cap-cover} on cap covering from Section~\ref{sec:macbeath}. Our proof is similar in spirit to the proofs of related covering lemmas by B\'{a}r\'{a}ny and Larman~\cite{BaL88}, B\'{a}r\'{a}ny~\cite{Bar00}, and Arya {\etal}~\cite{AFM17c}. Before proceeding with the proof, we recall some standard definitions. Throughout this section, we will use the term ``cap'' to mean a volume cap, that is, the nonempty intersection of $K$ with a halfspace $H$. Letting $h$ denote the hyperplane bounding $H$, the \emph{base} of the cap is $h \cap K$. Its \emph{width} is defined to be the distance between $h$ and the cap's opposing parallel supporting hyperplane (see Figure~\ref{fig:cap-cover}(a)). Given any cap $C$ of width $w$ and a real parameter $\lambda \geq 0$, we define its $\lambda$-expansion, denoted $C^{\lambda}$, to be the cap of $K$ cut by a hyperplane parallel to and at distance $\lambda w$ from this supporting hyperplane. Note that $C^{\lambda} = K$, if $\lambda w$ exceeds the width of $K$ along the defining direction.

%-----------------------------------------------------------------------
%-----------------------------------------------------------------------

Throughout this section, we assume that $K$ is a convex body that has been scaled to have unit volume. Let $\nu_0$ be a sufficiently small constant depending only on $d$. We begin by summarizing some known properties of Macbeath regions (defined in Section~\ref{sec:macbeath}). Claim~(i) is a variant of Lemma~1 of~\cite{ELR70} and was established by Br{\" o}nnimann, Chazelle, and Pach~\cite[Lemma~2.5]{BCP93}. Claim~(ii) is a straightforward adaptation of Lemma~2.8 in~\cite{AFM17c} and is based on ideas from~\cite[Lemma~2]{ELR70} and~\cite[Lemma~2.6]{BCP93}. Claim~(iii) is an immediate consequence of the definition of Macbeath regions.

%-----------------------------------------------------------------------
\begin{lem} \label{lem:mac-props}
Given a convex body $K$ in $\RE^d$ and a cap $C$ of $K$:
\begin{enumerate} \setlength{\itemsep}{-0.5ex}\setlength{\parsep}{0pt}%
\item[$(i)$] For $x, y \in K$, if $M^{1/5}(x) \cap M^{1/5}(y) \neq \emptyset$, then $M^{1/5}(x) \subseteq M(y)$.

\item[$(ii)$] If $C$ has volume at most $\nu_0$, then for any $\lambda \geq 1$, $C^{\lambda} \subseteq M^{3 d (2\lambda - 1)}(x)$, where $x$ is the centroid of the base of $C$.

\item[$(iii)$] For any $x \in C$, $M(x) \subseteq C^2$.
\end{enumerate}
\end{lem}
%-----------------------------------------------------------------------

Our proof of Lemma~\ref{lem:cap-cover} is based on the following lemma, which establishes a relationship between scalings of Macbeath regions and caps (see Figure~\ref{fig:cap-cover}(b)).

\begin{lem} \label{lem:mac-prop5}
Let $C$ be a cap of $K$ with volume at most $\nu_0$. Let $x$ denote the centroid of the base of the cap $C^{1/2}$. For any point $y \in K$ such that $M^{1/5}(x) \cap M^{1/5}(y) \neq \emptyset$, 
\[
    M^{1/5}(y) 
        ~ \subseteq ~ C
        ~ \subseteq ~ M^{45 d}(y).
\]
\end{lem}

\begin{proof}
Since $M^{1/5}(x) \cap M^{1/5}(y) \neq \emptyset$, Lemma~\ref{lem:mac-props}(i) implies that $M^{1/5}(y) \subseteq M(x)$ and $M^{1/5}(x) \subseteq M(y)$. Also, by Lemma~\ref{lem:mac-props}(iii), $M(x) \subseteq (C^{1/2})^2 = C$. Combining these, we obtain the following:
\[
    M^{1/5}(y) 
        ~ \subseteq ~ M(x) \subseteq C,
\]
which establishes the first inclusion. 

To prove the second inclusion, we apply Lemma~\ref{lem:mac-props}(ii), setting $C$ to $C^{1/2}$ and $\lambda$ to $2$, which yields 
\[
    C 
        ~ =         ~ (C^{1/2})^2 
        ~ \subseteq ~ M^{3 d(2\lambda - 1)}(x) 
        ~ =         ~ M^{9 d}(x).
\]
Recall that $M^{1/5}(x) \subseteq M(y)$. Scaling both of these centrally symmetric bodies by any positive factor about their respective centers preserves the inclusion (see, e.g., B{\'a}r{\'a}ny~\cite{Bar89}), and hence,
\[
    M^{9 d}(x) 
        ~ =         ~ (M^{1/5}(x))^{45 d} 
        ~ \subseteq ~ M^{45 d}(y).
\]
Putting these together, we obtain $C \subseteq M^{9 d}(x) \subseteq M^{45 d}(y)$, as desired.
\end{proof}

We are now ready to present the proof of Lemma~\ref{lem:cap-cover}. Before considering the general case, let us assume that all the caps in $\mathcal{C}$ have volume at most $\nu_0$. Let $\mathcal{M}$ be any maximal set of disjoint Macbeath regions of the form $M^{1/5}(x)$, where $x$ is the centroid of the base of the cap $C^{1/2}$ for some $C \in \mathcal{C}$. For each Macbeath region $M^{1/5}(x)$, define its expanded body $M'$ to be $M^{45 d}(x)$. We will show that $\mathcal{M}$ and $\mathcal{M'}$ satisfy the properties given in the lemma. Property~(i) is straightforward since $M'$ is related to $M$ by a constant scaling factor of $45 \SP d / (1/5) = 225 \SP d$. To show Property~(ii), consider any cap $C \in \mathcal{C}$. Let $x$ denote the centroid of the base of $C^{1/2}$. By the maximality of $\mathcal{M}$, there is a Macbeath region $M^{1/5}(y) \in \mathcal{M}$ such that $M^{1/5}(x) \cap M^{1/5}(y) \neq \emptyset$. Applying Lemma~\ref{lem:mac-prop5}, it follows that $M^{1/5}(y) \subseteq C \subseteq M^{45 d}(y)$. This establishes Property~(ii) and thus proves the lemma for the special case where all caps have volume at most $\nu_0$.

We now discuss the modifications required for handling the general case. For each cap $C \in \mathcal{C}$ whose volume exceeds $\nu_0$, we replace it by the cap $C^{\lambda}$, where $\lambda < 1$ is chosen so that the volume of $C^{\lambda}$ is exactly $\nu_0$. Otherwise, we retain the original cap $C$. Let $\mathcal{C'}$ represent the resulting set of caps. We construct the sets $\mathcal{M}$ and $\mathcal{M'}$ for the set $\mathcal{C'}$ exactly as described in the special case above. Finally, for each expanded body $M' \in \mathcal{M'}$, if its volume is at least $\nu_0$, we replace it by the convex body $K$. Otherwise, we retain the same body $M'$. Let $\mathcal{M''}$ denote the resulting set of expanded bodies. We claim that the sets $\mathcal{M}$ and $\mathcal{M''}$ satisfy the properties given in the lemma for the set $\mathcal{C}$. 

First, note that the argument given for the special case implies that the sets $\mathcal{M}$ and $\mathcal{M'}$ satisfy these properties for $\mathcal{C'}$. Since we replace the expanded body only if its volume is at least constant $\nu_0$, it follows that Property~(ii) holds for $\mathcal{M}$ and $\mathcal{M''}$ (the ratio of the volume of $M$'s expanded body to the volume of $M$ increases by a factor of at most $1/\nu_0$). 

To establish Property~(ii), consider a cap $C \in \mathcal{C}$. If the volume of $C$ is at most $\nu_0$, then the argument given in the special case shows that there exists a body $M \in \mathcal{M}$ such that $M \subseteq C \subseteq M'$, where $M'$ is $M$'s expanded body in $\mathcal{M'}$. Regardless of whether $M'$ is retained or replaced by $K$ in the construction of $\mathcal{M''}$, this property continues to hold. On the other hand, if the volume of $C$ exceeds $\nu_0$, then recall that it is replaced by a cap $C^{\lambda}$, where $\lambda < 1$ is chosen so that the volume of $C^{\lambda}$ is exactly $\nu_0$. The argument given in the special case shows that there exists a body $M \in \mathcal{M}$ such that $M \subseteq C^{\lambda} \subseteq M'$, where $M'$ is $M$'s expanded body in $\mathcal{M'}$. It follows that $\vol(M') \geq \vol(C^{\lambda}) = \nu_0$. Thus, $M'$ must be replaced by $K$ in constructing $\mathcal{M''}$. In other words, $M$'s expanded body in $\mathcal{M''}$ is $K$. Clearly, Property~(ii) holds since $M \subseteq C \subseteq K$. This completes the proof of Lemma~\ref{lem:cap-cover}.

%=======================================================================
\section{Concluding Remarks} \label{sec:conc}
%=======================================================================

In this paper, we have proved the existence of an $\eps$-approximation to a convex body $K$ in $\RE^d$, whose size is sensitive to the shape of the body expressed in terms of its area radius, $\arad(K)$. Our result yields a uniform bound, which means that the result holds for all $\eps \leq \eps_0$, where $\eps_0$ does not depend on $K$. We have shown that this bound is tight up to constant factors (depending on the dimension) as a function of the area radius. A notable feature of our approach is that it reduces the problem of approximating a convex body in $\RE^d$ to that of approximating a constant number of convex functions on $\RE^{d-1}$. The connection between the approximation of convex bodies and the approximation of convex functions has been observed elsewhere (see, e.g.,~\cite{Rot92, AAFM19, AFM17a, HaK15}). The techniques developed here may be of interest to future applications.

Our results require that the minimum width of $K$ be at least $\eps$. This width requirement seems to be necessary. Consider, for example, a $(d-2)$-dimensional unit ball $B$ embedded within $\RE^d$, and let $B_{\delta}$ denote its Minkowski sum with the $d$-dimensional Euclidean ball of radius $\delta \ll \eps$. By the optimality of Dudley's bound for Euclidean balls, $\Omega\big(1/\eps^{(d-3)/2}\big)$ facets are needed to approximate $B$, and hence this bound applies to $B_{\delta}$ as well. However, the surface area of $B_{\delta}$ can be made arbitrarily small as a function of $\delta$. Of course, the width condition can always be satisfied by first taking the Minkowski sum of $K$ with the Euclidean ball of radius $\eps/2$.

Several additional interesting questions are raised by our work. The area radius (or alternatively the $(d-1)$st intrinsic volume) is only one way to define a measure of shape sensitivity. In another work, we have demonstrated a bound based on the volume radius $\vrad(K)$, which is asymptotically superior to the bound presented here~\cite{ArM25}. The area-sensitive approach presented here may be applicable in contexts where the volume-sensitive approach is not. An example is that of approximating convex surface patches. The ultimate goal would be a construction that yields the polytope of minimum combinatorial complexity that approximates a given body. Unfortunately, existing hardness results suggest that this may not be solvable in polynomial time~\cite{DaJ90}.

%=======================================================================
\section{Acknowledgments} \label{sec:ack}
%=======================================================================

We thank Quentin Merig\'{o}t for pointing out the relationship to Gruber's result in Section~\ref{sec:nonuniform}. We also express our gratitude to the anonymous reviewers of an earlier draft of this paper for their extensive suggestions regarding the paper's structure, notation, and approach.

%=======================================================================
% References
%=======================================================================

\pdfbookmark[1]{References}{s:ref}
\bibliographystyle{plainurl}
\bibliography{shortcuts,convex}

\begin{thebibliography}{10}

\bibitem{AAFM19}
A.~Abdelkader, S.~Arya, G.~D. da~Fonseca, and D.~M. Mount.
\newblock Approximate nearest neighbor searching with non-{Euclidean} and
  weighted distances.
\newblock In {\em Proc.\ 30th Annu.\ ACM-SIAM Sympos.\ Discrete Algorithms},
  pages 355--372, 2019.
\newblock \href {https://doi.org/10.1137/1.9781611975482.23}
  {\path{doi:10.1137/1.9781611975482.23}}.

\bibitem{AHW87}
N.~Alon, D.~Haussler, and E.~Welzl.
\newblock Partitioning and geometric embedding of range spaces of finite
  {Vapnik-Chervonenkis} dimension.
\newblock In {\em Proc.\ Third Annu.\ Sympos.\ Comput.\ Geom.}, pages 331--340,
  1987.
\newblock \href {https://doi.org/10.1145/41958.41994}
  {\path{doi:10.1145/41958.41994}}.

\bibitem{AAFM22}
R.~Arya, S.~Arya, G.~D. da~Fonseca, and D.~M. Mount.
\newblock Optimal bound on the combinatorial complexity of approximating
  polytopes.
\newblock {\em ACM Trans.\ Algorithms}, 18:1--29, 2022.
\newblock \href {https://doi.org/10.1145/3559106} {\path{doi:10.1145/3559106}}.

\bibitem{AFM12a}
S.~Arya, G.~D. da~Fonseca, and D.~M. Mount.
\newblock Polytope approximation and the {Mahler} volume.
\newblock In {\em Proc.\ 23rd Annu.\ ACM-SIAM Sympos.\ Discrete Algorithms},
  pages 29--42, 2012.
\newblock \href {https://doi.org/10.1137/1.9781611973099.3}
  {\path{doi:10.1137/1.9781611973099.3}}.

\bibitem{AFM17b}
S.~Arya, G.~D. da~Fonseca, and D.~M. Mount.
\newblock Near-optimal $\varepsilon$-kernel construction and related problems.
\newblock In {\em Proc.\ 33rd Internat.\ Sympos.\ Comput.\ Geom.}, pages
  10:1--15, 2017.
\newblock URL: \url{https://arxiv.org/abs/1703.10868}, \href
  {https://doi.org/10.4230/LIPIcs.SoCG.2017.10}
  {\path{doi:10.4230/LIPIcs.SoCG.2017.10}}.

\bibitem{AFM17c}
S.~Arya, G.~D. da~Fonseca, and D.~M. Mount.
\newblock On the combinatorial complexity of approximating polytopes.
\newblock {\em Discrete Comput.\ Geom.}, 58(4):849--870, 2017.
\newblock \href {https://doi.org/10.1007/s00454-016-9856-5}
  {\path{doi:10.1007/s00454-016-9856-5}}.

\bibitem{AFM17a}
S.~Arya, G.~D. da~Fonseca, and D.~M. Mount.
\newblock Optimal approximate polytope membership.
\newblock In {\em Proc.\ 28th Annu.\ ACM-SIAM Sympos.\ Discrete Algorithms},
  pages 270--288, 2017.
\newblock \href {https://doi.org/10.1137/1.9781611974782.18}
  {\path{doi:10.1137/1.9781611974782.18}}.

\bibitem{AFM24}
S.~Arya, G.~D. da~Fonseca, and D.~M. Mount.
\newblock Economical convex coverings and applications.
\newblock {\em SIAM J.\ Comput.}, 53(4):1002--1038, 2024.
\newblock \href {https://doi.org/10.1137/23M1568351}
  {\path{doi:10.1137/23M1568351}}.

\bibitem{AMM09b}
S.~Arya, T.~Malamatos, and D.~M. Mount.
\newblock The effect of corners on the complexity of approximate range
  searching.
\newblock {\em Discrete Comput.\ Geom.}, 41:398--443, 2009.
\newblock \href {https://doi.org/10.1007/s00454-009-9140-z}
  {\path{doi:10.1007/s00454-009-9140-z}}.

\bibitem{ArM25}
S.~Arya and D.~M. Mount.
\newblock Optimal volume-sensitive bounds for polytope approximation.
\newblock {\em Discrete Comput.\ Geom.}, 74(4):839--871, 2025.
\newblock \href {https://doi.org/10.1007/s00454-025-00780-z}
  {\path{doi:10.1007/s00454-025-00780-z}}.

\bibitem{AMX12}
S.~Arya, D.~M. Mount, and J.~Xia.
\newblock Tight lower bounds for halfspace range searching.
\newblock {\em Discrete Comput.\ Geom.}, 47:711--730, 2012.
\newblock \href {https://doi.org/10.1007/s00454-012-9412-x}
  {\path{doi:10.1007/s00454-012-9412-x}}.

\bibitem{Bor00}
K.~{B{\" o}r{\" o}czky Jr.}
\newblock Approximation of general smooth convex bodies.
\newblock {\em Adv.\ Math.}, 153:325--341, 2000.
\newblock \href {https://doi.org/10.1006/aima.1999.1904}
  {\path{doi:10.1006/aima.1999.1904}}.

\bibitem{Bal91}
K.~Ball.
\newblock Volume ratios and a reverse isoperimetric inequality.
\newblock {\em J. London Math. Soc.}, 44(2):351--359, 1991.
\newblock \href {https://doi.org/10.1112/jlms/s2-44.2.351}
  {\path{doi:10.1112/jlms/s2-44.2.351}}.

\bibitem{Bar89}
I.~B{\'a}r{\'a}ny.
\newblock Intrinsic volumes and $f$-vectors of random polytopes.
\newblock {\em Math.\ Ann.}, 285:671--699, 1989.

\bibitem{Bar00}
I.~B{\'a}r{\'a}ny.
\newblock The technique of {M}-regions and cap-coverings: {A} survey.
\newblock {\em Rend.\ Circ.\ Mat.\ Palermo}, 65:21--38, 2000.
\newblock URL: \url{https://users.renyi.hu/~barany/}.

\bibitem{Bar07}
I.~B{\'a}r{\'a}ny.
\newblock Random polytopes, convex bodies, and approximation.
\newblock In W.~Weil, editor, {\em Stochastic Geometry}, volume 1892 of {\em
  Lecture Notes in Mathematics}, pages 77--118. Springer, 2007.
\newblock \href {https://doi.org/10.1007/978-3-540-38175-4_2}
  {\path{doi:10.1007/978-3-540-38175-4_2}}.

\bibitem{BaL88}
I.~B{\'a}r{\'a}ny and D.~G. Larman.
\newblock Convex bodies, economic cap coverings, random polytopes.
\newblock {\em Mathematika}, 35:274--291, 1988.
\newblock \href {https://doi.org/10.1112/S0025579300015266}
  {\path{doi:10.1112/S0025579300015266}}.

\bibitem{BeH92}
U.~Betke and M.~Henk.
\newblock Estimating sizes of a convex body by successive diameters and widths.
\newblock {\em Mathematika}, 39(2):247--257, 1992.
\newblock \href {https://doi.org/10.1112/S0025579300014984}
  {\path{doi:10.1112/S0025579300014984}}.

\bibitem{Bon18}
G.~Bonnet.
\newblock Polytopal approximation of elongated convex bodies.
\newblock {\em Advances in Geometry}, 18:105--114, 2018.
\newblock \href {https://doi.org/10.1515/advgeom-2017-0038}
  {\path{doi:10.1515/advgeom-2017-0038}}.

\bibitem{BoM87}
J.~Bourgain and V.~D. Milman.
\newblock New volume ratio properties for convex symmetric bodies.
\newblock {\em Invent.\ Math.}, 88:319--340, 1987.
\newblock \href {https://doi.org/10.1007/BF01388911}
  {\path{doi:10.1007/BF01388911}}.

\bibitem{BCP93}
H.~Br{\"o}nnimann, B.~Chazelle, and J.~Pach.
\newblock How hard is halfspace range searching?
\newblock {\em Discrete Comput.\ Geom.}, 10:143--155, 1993.
\newblock \href {https://doi.org/10.1007/BF02573971}
  {\path{doi:10.1007/BF02573971}}.

\bibitem{BrG95}
H.~Br{\" o}nnimann and M.~T. Goodrich.
\newblock Almost optimal set covers in finite {VC}-dimension,.
\newblock {\em Discrete Comput.\ Geom.}, 14:463--479, 1995.
\newblock \href {https://doi.org/10.1007/BF02570718}
  {\path{doi:10.1007/BF02570718}}.

\bibitem{BrI76}
E.~M. Bronshteyn and L.~D. Ivanov.
\newblock The approximation of convex sets by polyhedra.
\newblock {\em Siberian Math.\ J.}, 16:852--853, 1976.
\newblock \href {https://doi.org/10.1007/BF00967115}
  {\path{doi:10.1007/BF00967115}}.

\bibitem{Bro08}
E.~M. Bronstein.
\newblock Approximation of convex sets by polytopes.
\newblock {\em J.\ Math.\ Sci.}, 153(6):727--762, 2008.
\newblock \href {https://doi.org/10.1007/s10958-008-9144-x}
  {\path{doi:10.1007/s10958-008-9144-x}}.

\bibitem{Cla93}
K.~L. Clarkson.
\newblock Algorithms for polytope covering and approximation.
\newblock In {\em Proc.\ Third Internat.\ Workshop Algorithms Data Struct.},
  pages 246--252, 1993.

\bibitem{Cla06}
K.~L. Clarkson.
\newblock Building triangulations using $\varepsilon$-nets.
\newblock In {\em Proc.\ 38th Annu.\ ACM Sympos.\ Theory Comput.}, pages
  326--335, 2006.
\newblock \href {https://doi.org/10.1145/1132516.1132564}
  {\path{doi:10.1145/1132516.1132564}}.

\bibitem{DaJ90}
G.~Das and D.~Joseph.
\newblock The complexity of minimum convex nested polyhedra.
\newblock In {\em Proc.\ Second Canad.\ Conf.\ Comput.\ Geom.}, pages 296--301,
  1990.

\bibitem{Car76}
M.~P. do~Carmo.
\newblock {\em Differential Geometry of Curves and Surfaces}.
\newblock Prentice Hall, 1976.

\bibitem{Dud74}
R.~M. Dudley.
\newblock Metric entropy of some classes of sets with differentiable
  boundaries.
\newblock {\em J.\ Approx.\ Theory}, 10(3):227--236, 1974.
\newblock \href {https://doi.org/10.1016/0021-9045(74)90120-8}
  {\path{doi:10.1016/0021-9045(74)90120-8}}.

\bibitem{DGJ19}
K.~Dutta, A.~Ghosh, B.~Jartoux, and N.~H. Mustafa.
\newblock Shallow packings, semialgebraic set systems, {Macbeath} regions and
  polynomial partitioning.
\newblock {\em Discrete Comput.\ Geom.}, 61:756--777, 2019.
\newblock \href {https://doi.org/10.1007/s00454-019-00075-0}
  {\path{doi:10.1007/s00454-019-00075-0}}.

\bibitem{ELR70}
G.~Ewald, D.~G. Larman, and C.~A. Rogers.
\newblock The directions of the line segments and of the $r$-dimensional balls
  on the boundary of a convex body in {Euclidean} space.
\newblock {\em Mathematika}, 17:1--20, 1970.
\newblock \href {https://doi.org/10.1112/S0025579300002655}
  {\path{doi:10.1112/S0025579300002655}}.

\bibitem{Fed96}
H.~Federer.
\newblock {\em Geometric Measure Theory}.
\newblock Springer Berlin, Heidelberg, 1996.
\newblock \href {https://doi.org/10.1007/978-3-642-62010-2}
  {\path{doi:10.1007/978-3-642-62010-2}}.

\bibitem{Gru93a}
P.~M. Gruber.
\newblock Aspects of approximation of convex bodies.
\newblock In P.~M. Gruber and J.~M. Wills, editors, {\em Handbook of Convex
  Geometry}, chapter 1.10, pages 319--345. North-Holland, 1993.
\newblock \href {https://doi.org/10.1016/B978-0-444-89596-7.50015-8}
  {\path{doi:10.1016/B978-0-444-89596-7.50015-8}}.

\bibitem{HaK15}
S.~Har-Peled and N.~Kumar.
\newblock Approximating minimization diagrams and generalized proximity search.
\newblock {\em SIAM J.\ Comput.}, 44:944--974, 2015.
\newblock \href {https://doi.org/10.1137/140959067}
  {\path{doi:10.1137/140959067}}.

\bibitem{Joh48}
F.~John.
\newblock Extremum problems with inequalities as subsidiary conditions.
\newblock In {\em Studies and Essays Presented to R. Courant on his 60th
  Birthday}, pages 187--204. Interscience Publishers, Inc., New York, 1948.

\bibitem{Kup08}
G.~Kuperberg.
\newblock From the {Mahler} conjecture to {Gauss} linking integrals.
\newblock {\em Geom.\ Funct.\ Anal.}, 18:870--892, 2008.
\newblock \href {https://doi.org/10.1007/s00039-008-0669-4}
  {\path{doi:10.1007/s00039-008-0669-4}}.

\bibitem{Mac52}
A.~M. Macbeath.
\newblock A theorem on non-homogeneous lattices.
\newblock {\em Ann.\ of Math.}, 56:269--293, 1952.
\newblock \href {https://doi.org/10.2307/1969800} {\path{doi:10.2307/1969800}}.

\bibitem{Mat02}
J.~Matou{\v{s}}ek.
\newblock {\em Lectures on Discrete Geometry}.
\newblock Springer-Verlag, 2002.
\newblock \href {https://doi.org/10.1007/978-1-4613-0039-7}
  {\path{doi:10.1007/978-1-4613-0039-7}}.

\bibitem{McV75}
D.~E. McClure and R.~A. Vitalie.
\newblock Polygonal approximation of plane convex bodies.
\newblock {\em J.\ Math.\ Anal.\ Appl.}, 51:326--358, 1975.
\newblock \href {https://doi.org/10.1016/0022-247X(75)90125-0}
  {\path{doi:10.1016/0022-247X(75)90125-0}}.

\bibitem{McM75}
P.~McMullen.
\newblock Non-linear angle-sum relations for polyhedral cones and polytopes.
\newblock {\em Math.\ Proc.\ Cambridge Philos.\ Soc}, 78(2):247--261, 1975.
\newblock \href {https://doi.org/10.1017/S0305004100051665}
  {\path{doi:10.1017/S0305004100051665}}.

\bibitem{McM91}
P.~McMullen.
\newblock Inequalities between intrinsic volumes.
\newblock {\em Monatshefte f{\" u}r Mathematik}, 111:47--53, 1991.
\newblock \href {https://doi.org/10.1007/BF01299276}
  {\path{doi:10.1007/BF01299276}}.

\bibitem{Mus22}
N.~H. Mustafa.
\newblock {\em Sampling in Combinatorial and Geometric Set Systems}, volume 265
  of {\em Mathematical Surveys and Monographs}.
\newblock AMS, 2022.
\newblock \href {https://doi.org/10.1090/surv/265}
  {\path{doi:10.1090/surv/265}}.

\bibitem{MuR14}
N.~H. Mustafa and S.~Ray.
\newblock Near-optimal generalisations of a theorem of {Macbeath}.
\newblock In {\em Proc.\ 31st Internat.\ Sympos.\ on Theoret.\ Aspects of
  Comp.\ Sci.}, pages 578--589, 2014.
\newblock \href {https://doi.org/10.4230/LIPIcs.STACS.2014.578}
  {\path{doi:10.4230/LIPIcs.STACS.2014.578}}.

\bibitem{Naz12}
F.~Nazarov.
\newblock The {H\"o}rmander proof of the {Bourgain-Milman} theorem.
\newblock In {\em Geometric Aspects of Functional Analysis}, pages 335--343.
  Springer, 2012.
\newblock \href {https://doi.org/10.1007/978-3-642-29849-3_20}
  {\path{doi:10.1007/978-3-642-29849-3_20}}.

\bibitem{Roc67}
R.~T. Rockafellar.
\newblock Conjugates and {Legendre} transforms of convex functions.
\newblock {\em Canad.\ J.\ Math.}, pages 200--205, 1967.
\newblock \href {https://doi.org/10.4153/CJM-1967-012-4}
  {\path{doi:10.4153/CJM-1967-012-4}}.

\bibitem{Roc97}
R.~T. Rockafellar.
\newblock {\em Convex Analysis}.
\newblock Princeton University Press, Princeton, NJ, 1997.
\newblock \href {https://doi.org/10.1515/9781400873173}
  {\path{doi:10.1515/9781400873173}}.

\bibitem{Rot92}
G.~Rote.
\newblock The convergence rate of the sandwich algorithm for approximating
  convex functions.
\newblock {\em Computing}, 48:337--361, 1992.
\newblock \href {https://doi.org/10.1007/BF02238642}
  {\path{doi:10.1007/BF02238642}}.

\bibitem{Sch87}
R.~Schneider.
\newblock Polyhedral approximation of smooth convex bodies.
\newblock {\em J.\ Math.\ Anal.\ Appl.}, 128:470--474, 1987.
\newblock \href {https://doi.org/10.1016/0022-247X(87)90197-1}
  {\path{doi:10.1016/0022-247X(87)90197-1}}.

\bibitem{Tot48}
L.~F. Toth.
\newblock Approximation by polygons and polyhedra.
\newblock {\em Bull.\ Amer.\ Math.\ Soc.}, 54:431--438, 1948.
\newblock \href {https://doi.org/10.1090/S0002-9904-1948-09022-X}
  {\path{doi:10.1090/S0002-9904-1948-09022-X}}.

\end{thebibliography}

\end{document}